\documentclass[final,11pt]{article}

\usepackage{graphics,graphicx}
\usepackage{amsmath,amsthm,amssymb}
\usepackage[spanish, english]{babel}
\usepackage{textcomp}
\usepackage{rotating}
\usepackage{lscape}
\usepackage{longtable}
\usepackage{color}
\usepackage[latin1,utf8]{inputenc}
\usepackage[T1]{fontenc}
\usepackage{fancyhdr}
\usepackage{pst-all}
\usepackage[bookmarks = true, colorlinks=true, linkcolor = blue, citecolor = blue, menucolor = black, urlcolor = black, plainpages=false]{hyperref}
\usepackage[left=2.5cm,top=2.5cm,right=2.5cm,nohead,foot=1.5cm]{geometry}
\usepackage[T1]{fontenc}
\usepackage{rawfonts}
\usepackage{amsfonts}
\usepackage{enumerate}
\usepackage{pstricks}
\usepackage{layout}
\usepackage{colortbl}
\usepackage{multimedia}
\usepackage{times}
\usepackage{array}
\usepackage{supertabular}
\usepackage{xspace}
\usepackage{ragged2e}
\usepackage{multicol}
\usepackage[active]{srcltx}
\usepackage{makeidx}
\usepackage{dsfont}
\usepackage{bbm}
\usepackage{breakcites}
\usepackage{color}
\usepackage[toc,page]{appendix}
\usepackage[explicit]{titlesec}

\newtheorem{Theorem}{Theorem}[section]
\newtheorem{Definition}[Theorem]{Definition}

\newtheorem{Proposition}[Theorem]{Proposition}

\newtheorem{Lemma}[Theorem]{Lemma}
\newtheorem{Corollary}[Theorem]{Corollary}

\newtheorem{Remark}[]{Remark}
\newenvironment{Proof}[1][Proof]{{\bfseries\text{#1.\;\;}}}{\hfill $\square$}%

\newtheorem{Assum}{Assumption}

\numberwithin{equation}{section} 
\numberwithin{Theorem}{section}

\pdfoutput=1

\begin{document}





\title{\Large{A bootstrap-based method to estimate directional extreme risk regions at high levels}}
\author{Ra\'ul Torres (ratorres@est-econ.uc3m.es)\\
Elena Di Bernardino (elena.di\_bernardino@cnam.fr)\\
Henry Laniado (hlaniado@eafit.edu.co)\\
Rosa E. Lillo (rosaelvira.lillo@uc3m.es)}

\maketitle


\begin{abstract}
In multivariate extreme value theory (\textit{MEVT}), the focus is on analysis outside of the observable sampling zone, which implies that the region of interest is associated to high risk levels. This work provides tools to include directional notions into the \textit{MEVT}, giving the opportunity to characterize the recently introduced directional multivariate quantiles (\textit{DMQ}) at high levels. Then, an \emph{out-sample} estimation method for these quantiles is given. A bootstrap procedure carries out the estimation of the tuning parameter in this multivariate framework and helps with the estimation of the \textit{DMQ}. Asymptotic normality for the proposed estimator is provided and the methodology is illustrated with simulated data-sets. Finally, a real-life  application to a financial case is also performed.
\end{abstract}


\maketitle

\section{Introduction}\label{sec:intro}

The estimation of extreme level curves is important for identifying extreme events and for characterizing the joint tails of multidimensional distributions. They are usually considered as quantiles at high levels; that is, they are linked with a probability $\alpha$ of occurrence of certain event, where $\alpha$ is a very small number. This proposal considers a non-parametric approach for values of $\alpha$ lower or equal than $1/n$, where $n$ denotes the sample size. This  implies that the number of data points that fall beyond the quantile curve is small and can even be zero, thus we are outside of the observable sample zone, or in other words, in the \textit{out-sample} estimation framework. This lack of relevant data points makes the estimation difficult, making it necessary to introduce tools from the multivariate extreme value theory (\textit{MEVT}).

The main purpose of this paper is to provide an extension of the $MEVT$ by introducing the directional approach and also to give an \textit{out-sample} estimation method for the directional multivariate quantiles (\textit{DMQ}) introduced in \cite{torres1st} and \cite{torres2nd}. In these papers, the directional setting refers to the inclusion of a parameter of direction $\mathbf{u}$ that allows analysis of data by looking at the cloud of observations from different perspectives. Accurate assessments of these quantiles are sought in a diversity of applications from financial risk management (e.g. \cite{torres1st}) to environmental impact assessment (e.g. \cite{torres2nd}). A non-parametric estimation method was developed in \cite{torres1st} to estimate the directional quantile based on the empirical probability distribution, which is valid just for the \textit{in-sample} scenario; that is $\alpha > 1/n$.

Both scenarios, \textit{in-sample} and \textit{out-sample}, have been widely studied in the univariate setting and recently the literature has focused on the extension to the multivariate context. Some relevant references in this area can be grouped into three categories as follows. Firstly, estimation under optimization processes, for instance,  based on optimization over linear quantile regression (see, e.g., \cite{chaudhuri,hapasi,mukhopadhyay,mizera,girard}). In this case, an example of the \textit{in-sample} framework is given by \cite{chaudhuri} for geometric quantiles. \cite{girard}  also proposed an \textit{out-sample} estimation method for geometric quantiles.

A second category contains methods determining level curves of joint density functions in such a way that the set of points outside those contours has a probability equal to a given level $\alpha$. These methods easily describe inner and outer regions at the given level and inherently cover the infinite set of directions through those contours (e.g., \cite{cedh,edhk}). The estimators proposed in this category have been developed mainly for the \textit{out-sample} framework. For instance, \cite{cedh}  provided estimation of bivariate contour levels for some joint densities with elliptical and non-elliptical distributions, considering cases with asymptotic dependence and asymptotic independence. Other methodologies in this category are based on the trimming through depth functions. \cite{serfling2}  described \textit{in-sample} methods considering different depth functions and \cite{yihe} presented an \textit{out-sample} contour estimation based on the Tukey depth.

Last category considers level curve estimations using either joint distribution or survival functions (e.g. \cite{dehaan1995,fpsll,bcsll,chebana2009,dblmp}). Works based on copulas are also classified in this group (e.g. \cite{chebana2,durante,setal,binos}). These works have introduced the estimation procedures in both contexts, \textit{in-sample} and \textit{out-sample}, but most of them present the theory and related  applications only in the bivariate case. Since the proposal developed in this work is somehow based on distributions, it belongs to this third category.

As we have mentioned before, the methodology developed in this work includes a directional notion and we want to highlight its importance in our contribution. One can find in the literature a few references dealing with this notion. \cite{chaudhuri} is one of the first works that includes directions. However, this multivariate aspect starts to take importance just in the past decade, where an accurate assessment of risk regions arises in a diversity of applications. For instance, \cite{ep} studied bounds for multivariate financial risks, highlighting the usefulness of analysis considering two particular directions. \cite{bcsll} presented a bivariate quantile application to air quality where the directions are related to the four classical orthants.

Other examples describing the importance of directions are \cite{hapasi}, where it was proposed directional projections to show a relationship between their quantile trimming and the trimming obtained through the Tukey depth. \cite{mizera} used a similar idea to build multivariate growth chart applications. \cite{pateiro} provided a directional projection-based definition for infinite-dimensional multivariate quantiles in Hilbert spaces. In financial risk management \cite{torres1st} showed the advantage of using the portfolio weights of investment as the direction of analysis to provide an upper bound for the maximum loss.  \cite{torres2nd} performed an application to environmental impact assessment, where it can be seen the improvement of identifying extremes by using the first principal component as a direction of analysis.

Therefore, inspired by \cite{dehaan1995} where an \textit{out-sample} estimator for bivariate level curves of a distribution function $F$ was established, the contribution in this paper is threefold: 1) to include the directional framework given in \cite{torres1st,torres2nd} in the \textit{MEVT} analysis, 2) to establish an estimator those directional high level quantiles in a general dimension $d$ and 3) to provide a non-parametric estimation method for these high level directional quantiles and some asymptotic results.

The paper is organized as follows. In Section \ref{sec:dirExtremes} we summarize the main definitions and results related to the directional framework used in the paper. Section \ref{sec:assum} introduces definitions from the multivariate extreme theory to fix conditions over the random vector $\mathbf{X}$ that allow to ensure the results under the directional framework. In Section \ref{sec:qCharacterization}, we describe the characterization of the elements of the \textit{DMQ} at high levels, based on the heuristic ideas in \cite{dehaan1995}. Section \ref{sec:inference} develops statistical tools to perform an \textit{out-sample} estimation of the \textit{DMQ}. The high level estimator is introduced in Section \ref{subsec:qEstimator}. Asymptotic normality of this estimator is presented in Section \ref{subsec:normality}. Later on, we adapt a bootstrap-based method to deal with the tuning parameter. Section \ref{sec:tExample} illustrates the performance of our  multivariate estimation procedure (in dimensions $d=2$ and $d=3$) in a multivariate $t-$distribution case. Section \ref{sec:realCase} presents a directional analysis over daily filtered returns of three different international indices. Finally, in Section \ref{sec:conclusions} some conclusions and perspectives are provided. Proofs and auxiliary results are postponed to  Appendix \ref{sec:proofs}.

\section{Directional multivariate quantiles}\label{sec:dirExtremes}
This section introduces the preliminary definitions and notation necessary to understand the contributions of the paper. Our directional multivariate setting is based on the work developed in \cite{laniadoIME}. Firstly, we recall  the notion of oriented orthant.
\begin{Definition}[Oriented orthant]\label{def:orthant}
An oriented orthant in $\mathbb{R}^{d}$ with vertex $\mathbf{x}$ in direction $\mathbf{u}$ is defined by,
\begin{equation*}\label{eq:convexcone}
	\mathfrak{C}_{\mathbf{x}}^{R_{\mathbf{u}}}=\{\mathbf{z}\in \mathbb{R}^{d}:R_{\mathbf{u}}(\mathbf{z}-\mathbf{x})\geq \bf{0}\},
\end{equation*}
where $\mathbf{u}\in\{\mathbf{z}\in\mathbb{R}^{d}:||\mathbf{z}||=1\}$ and $R_{\mathbf{u}}$ is an orthogonal matrix such that $R_{\mathbf{u}}\mathbf{u}=\mathbf{e}$, with $\mathbf{e}=\frac{1}{\sqrt{d}}(1,\ldots,1)'$.
\end{Definition}
Note that an oriented orthant is a translation by $\mathbf{x}$ and rotation by $R_{\mathbf{u}}$ of the non-negative Euclidean
orthant. \cite{torres1st} pointed out that $R_{\mathbf{u}}$ is not unique for $d\geq 3$. Hence, in order to guarantee uniqueness, they stated the following. Let $\mathbf{u}$ be a unit vector with non-null components and let $M_{\mathbf{u}}$ and $M_{\mathbf{e}}$ be matrices defined as,
\begin{equation*}\label{eq:basicMat}
M_{\mathbf{u}} = [\mathbf{u},\ \ sgn(u_{2})\mathbf{e}_{2},\ \ \cdots,\ \ sgn(u_{d})\mathbf{e}_{d}]\quad\text{ and }\quad M_{\mathbf{e}} = [\mathbf{e},\ \ \mathbf{e}_{2},\ \ \cdots,\ \ \mathbf{e}_{d}],
\end{equation*}
where $u_{i}$, $i=1,\ldots,d$ is the $i-$th component of $\mathbf{u}$, $sgn(\cdot)$ is the scalar sign function and $\mathbf{e}_{i}$ is the $i-$th column of the $d\times d$ identity matrix. Then $M_{\mathbf{u}}$ and $M_{\mathbf{e}}$ have full rank and unique QR decomposition,
\[M_{\mathbf{u}} = Q_{\mathbf{u}}T_{\mathbf{u}}\quad\text{ and }\quad M_{\mathbf{e}} = Q_{\mathbf{e}}T_{\mathbf{e}},\]
such that $T_{\mathbf{u}}$, $T_{\mathbf{e}}$ are triangular matrices with positive diagonal elements and $Q_{\mathbf{u}}$, $Q_{\mathbf{e}}$ are the correspondent orthogonal matrices (see \cite{horn}[Theorem 2.1.14, p.g. 89]). Therefore, \cite{torres1st}   defined the QR oriented orthant as follows.
\begin{Definition}[QR oriented orthant]\label{def:orthantQR}
The QR oriented orthant with vertex $\mathbf{x}$ in direction $\mathbf{u}$, denoted as $\mathfrak{C}_{\mathbf{x}}^{\mathbf{u}}$, is the oriented orthant satisfying $R_{\mathbf{u}}=Q_{\mathbf{e}}Q'_{\mathbf{u}}$.
\end{Definition}

\begin{figure}[htbp]\centering
\includegraphics[height=4.5cm,width=5cm]{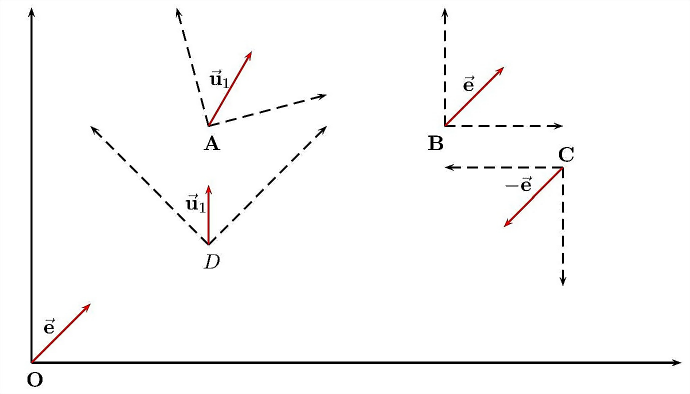}
\includegraphics[height=4.5cm,width=5cm]{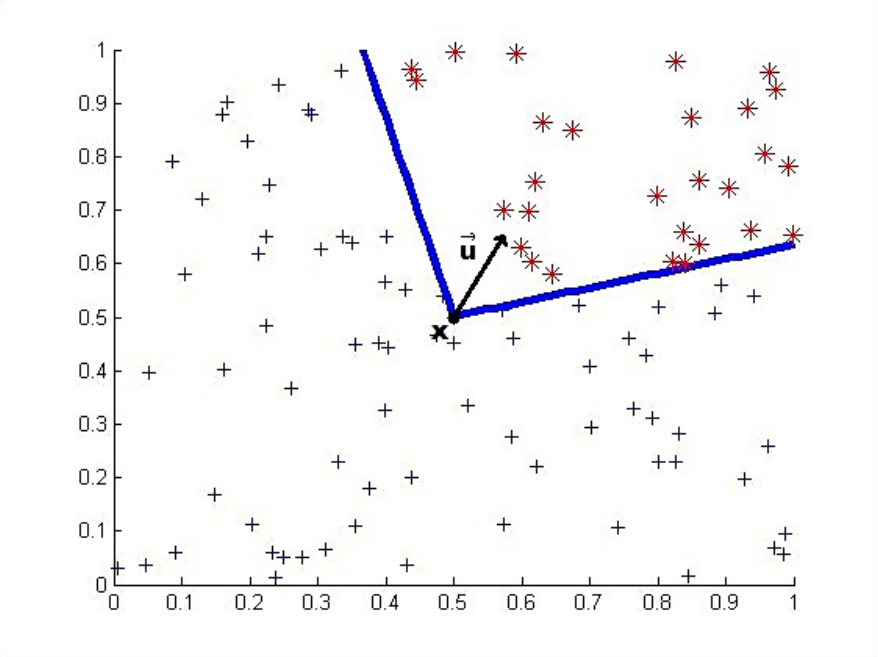}
\includegraphics[height=4.5cm,width=5cm]{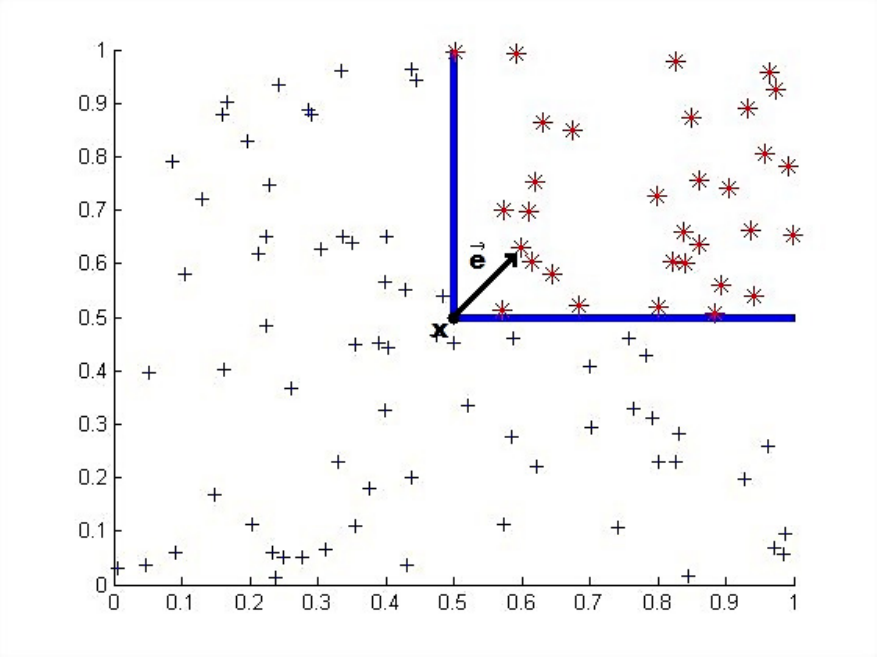}\vspace{-0.25cm}
\caption{Examples of QR oriented orthants for different vertexes $\{O,A,B,C,D,x\}$ and different direction $\{\mathbf{e},-\mathbf{e},\mathbf{u}_{1},\mathbf{u}_{2},\mathbf{u}\}$.}\label{fig:qrOrthant}
\end{figure}

Figure \ref{fig:qrOrthant} illustrates the QR oriented orthant in Definition \ref{def:orthantQR} for different vertexes and  directions. One can observe that the mass accumulated inside a QR oriented orthant with vertex $\mathbf{x}$ and direction $\mathbf{u}$ corresponds to the probability of such orthant and moreover,  it is equivalent to evaluate the vertex $\mathbf{x}$ in the survival function of the rotated random vector $R_{\mathbf{u}}\mathbf{X}$. We remark that directions with non-null components is not a restrictive assumption, because if there exists a direction $\mathbf{u}$ with zero in one or more components, this implies independence of the corresponding marginals from the joint extreme behavior of interest.

Also, quantiles at certain level $\alpha$ in direction $\mathbf{u}$ have been defined in \cite{torres1st} as follows.
\begin{Definition}[Directional multivariate quantile (\textit{DMQ})]\label{def:dirQ}
Let $\mathbf{X}$ be a random vector with associated probability distribution $\mathbb{P}$. Then the directional multivariate quantile at level $\alpha$ in direction $\mathbf{u}$ is defined as
\begin{equation}\label{eq:qExt}
	\mathcal{Q}_{\mathbf{X}}(\alpha,\mathbf{u}):=\partial\{\mathbf{x}\in\mathbb{R}^{d}:\mathbb{P}(\mathfrak{C}_{\mathbf{x}}^{-\mathbf{u}})\geq 1-\alpha\},
\end{equation}
where $\partial$ denotes the boundary of the considered  set  and $0\leq\alpha\leq 1$.
\end{Definition}

Examples of the previous definition applied on a multivariate $t-$distribution are given for high levels of $\alpha$ in different directions in Figure \ref{fig:exampleT} (dimension $d=2$) and Figure \ref{fig:qTeo3D} (dimension $d=3$). In the univariate setting, extremes are analyzed considering the two possibilities of exceeding from either distributions or survival functions and most of the extensions of these analyses to the multivariate setting have also been concentrated on these two types of exceeding. The interested reader is referred to \cite{shiau,salvadori,ep} for extensions on the bivariate case and also to \cite{gupta,bernardino,bernardino3} for some generalized multivariate versions.

However, the multivariate setting offers infinite possibilities of exceeding to be considered and our directional framework explores these alternatives.  First, note that Definition \ref{def:dirQ} provides a multivariate quantile based on two free parameters: traditional $\alpha-$level and a unit vector of direction $\mathbf{u}$. Given the \textit{out-sample} framework of this paper, the $\alpha-$level will be considered lower or equal than $1/n$ for a sample size $n$, but more important is to highlight that the parameter of direction is completely free and each choice implies different hyper-curves, each of those focusing on the extreme behavior of $\mathbf{X}$ in particular areas of its support. Note that the possibilities based on the distributions and the survival functions of a random vector $\mathbf{X}$ are provided through the directions $\mathbf{e}$, $-\mathbf{e}$ respectively. Hereafter we will call these directions: \textit{classical directions}.

Second, we point out that there are other interesting directions to be taken into consideration. For instance in portfolio optimization, the direction given by the portfolio weights of investments is of particular interest because it takes into account the losses due to the composition of the investment in a portfolio (see \cite{laniadoIME,torres1st}). In environmental phenomena, the directional approach has also been applied to detect extreme events by considering the direction of maximum variability of the data (see \cite{torres2nd}). In summary, different contexts or phenomena could suggest different particular directions either by external information or other causes, so the direction is a parameter to be chosen by a practitioner helping to capture the overall behavior of the data and to improve the visualization of results.

%

\section{Directional \textit{MEVT}: A probabilistic approach}\label{sec:assum}

In this section, we introduce conditions over $\mathbf{X}$ in order to extend classic \textit{MEVT} by using the directional framework  in Section \ref{sec:dirExtremes}. This allows us to provide a general theory that includes a free parameter of direction $\mathbf{u}$. Those conditions involve notions of multivariate regular variation. The interested reader is referred to \cite{resnick1, deHaan-ferreira, resnick2}.

\begin{Assum}\label{assum2}
The support of $\mathbf{X}$ is all $\mathbb{R}^{d}$.
\end{Assum}

This assumption is introduced to guarantee that, independently of the chosen direction $\mathbf{u}$, $R_{\mathbf{u}}\, \mathbf{X}$ possesses a part of its support on the positive orthant. Indeed if $\mathbf{X}$ has a bounded support, one should be aware of the fact that many directions  become uninformative in terms of maximums. Then, if Assumption \textbf{\ref{assum2}} is not satisfied, it is advisable to fix first the desired direction $\mathbf{u}$ and then to make the corresponding analysis in terms of classical \textit{MEVT} for the vector $R_{\mathbf{u}}\, \mathbf{X}$.

We now characterize the  right tail behavior of $\mathbf{X}$ (see Assumption \textbf{\ref{assum3}}) and later we show that this characterization is inherited for any rotation $R_{\mathbf{u}}$ of the random vector (see Proposition \ref{prop:rotRegVar}).

\begin{Definition}[First order multivariate regular variation]\label{def:regVarying}
A random vector $\mathbf{X}$ has first order multivariate regular variation with tail index $\gamma$, if there exists a real-value function $\phi(t)>0$ that is regularly varying at infinity\footnote{A function $\phi(\cdot)\in RV_{1/\gamma}$, if it holds that $\lim_{x\rightarrow\infty}\frac{\phi(tx)}{\phi(x)}=t^{\frac{1}{\gamma}}$, for all $t>0$.} with exponent $1/\gamma$, denoted by $RV_{1/\gamma}$, and a non-zero measure $\mu(\cdot)$ on the Borel $\sigma-$field $[-\infty,\, \infty]^{d}\backslash\{\mathbf{0}\}$, such that, if $t\rightarrow\infty$,
\begin{equation}\label{eq:regVarying}
t\,\mathbb{P}\left[\frac{\mathbf{X}}{\phi(t)}\,\in \, \cdot\right]\stackrel{v}{\rightarrow} \,\mu(\cdot),
\end{equation}
where $\stackrel{v}{\rightarrow}$ means vague convergence (see, e.g., \cite{jessen,resnick1}).
\end{Definition}
If $\mathbf{X}$ verifies Definition \ref{def:regVarying} with $\gamma > 0$, then it is called a heavy tailed random vector and the measure of convergence $\mu(\cdot)$ in \eqref{eq:regVarying} has the homogeneity property of order $\gamma$, i.e., $\mu(cB)=c^{-\gamma}\mu(B)$, for all $c>0$ and every Borel set $B$.

\begin{Assum}\label{assum3}
$\mathbf{X}$ has first order multivariate regular variation with tail index $\gamma > 0$.
\end{Assum}

Note that Assumption \textbf{\ref{assum3}} is a stronger condition than $\mathbf{X}$ belonging to a multivariate max-domain of attraction (see \cite{deHaan-ferreira}). Indeed, as it is mentioned in \cite{resnick1}[Remark 6.1], \textbf{\ref{assum3}} implies tail equivalence among marginal distributions, reducing all marginal tail indexes to $\gamma$. In practice, tails could be different and in this case, there are some techniques to overcome this flaw. (see for instance \cite{resnick1}[Section 6.5.6]). We refer to Remark \ref{remarkAssum} below for further comments on this point.

\begin{Proposition}\label{prop:rotRegVar}
If $\mathbf{X}$ satisfies Assumption \textbf{\ref{assum3}}  with tail index $\gamma$, then the random vector $Q\mathbf{X}$ has first order multivariate regular variation with tail index $\gamma$, for any orthogonal transformation $Q$.
\end{Proposition}
Proposition \ref{prop:rotRegVar} is a special case of Proposition A.1 in \cite{basrak_et_al} for a deterministic matrix $Q$. An alternative proof of Proposition \ref{prop:rotRegVar} is given in  Appendix \ref{sec:proofs}.

\begin{Definition}[Second order multivariate regular variation]\label{def:2ndOrderRegVarying}
A random vector $\mathbf{X}$ has second order multivariate regular variation with indexes $(\gamma,\, \pi)$, if there exist functions $\phi(\cdot)\in RV_{1/\gamma}$ and $\Lambda(t)\rightarrow 0$, such that $|\Lambda|\in RV_{\pi}$, $\pi\leq 0$; satisfying,
\begin{equation*}\label{eq:2ndOrder}
\frac{t\mathbb{P}\left[\frac{\mathbf{X}}{\phi(t)}\in B\right]-\mu(B)}{\Lambda(\phi(t))}\rightarrow \psi(B)\, <\infty,
\end{equation*}
locally uniformly for all relatively compact rectangles $B\in[-\infty,\, \infty]^{d}\backslash\{\mathbf{0}\}$, where $\psi(\cdot)$ is not identically zero. (see \cite{resnickHidden}).
\end{Definition}

\begin{Assum}\label{assum4}
$\mathbf{X}$ has second order multivariate regular variation with indexes $(\gamma,\, \pi)$.
\end{Assum}

\begin{Proposition}\label{prop:rotRegVar2}
If $\mathbf{X}$ satisfies Assumption \textbf{\ref{assum4}}  with indexes $(\gamma,\, \pi)$, then the random vector $Q\mathbf{X}$ has second order regular variation with indexes $(\gamma,\, \pi)$, for any orthogonal transformation $Q$.
\end{Proposition}
Proof of Proposition \ref{prop:rotRegVar2} is given in  Appendix \ref{sec:proofs}. Assumption \textbf{\ref{assum4}} will be crucial in Section \ref{subsec:normality} to prove asymptotic normality for the proposed estimator of the \textit{DMQ}. From now on, we consider that a random vector $\mathbf{X}$ satisfies Assumptions \textbf{\ref{assum2}-\ref{assum4}}.

\subsection{Characterization of the \textit{DMQ} at high levels}\label{sec:qCharacterization}

The aim of this paragraph is to characterize the points belonging to $\mathcal{Q}_{\mathbf{X}}(\alpha,\mathbf{u})$ in Definition \ref{def:dirQ} for small values of the $\alpha$ level, ($\alpha \leq 1/n$). Our proposal is based in two main aspects. The first one is the \textit{quasi-orthogonal invariace property} given in \cite{torres1st}[Property 3.8], i.e.,
\begin{equation}\label{eq:ortProp}
\mathcal{Q}_{\mathbf{X}}(\alpha,\mathbf{u}) = R_{\mathbf{u}}'\mathcal{Q}_{R_{\mathbf{u}}\mathbf{X}}(\alpha,\mathbf{e}),
\end{equation}
and the second one is referred to the heuristic ideas of the bivariate quantile parameterization given in \cite{dehaan1995} extended to a general directional multivariate context.

Let $F_{\mathbf{u}}$ be the distribution function of the random vector $R_{\mathbf{u}}{X}$. Note that $\mathcal{Q}_{R_{\mathbf{u}}\mathbf{X}}(\alpha,\mathbf{e})$ is the set of points such that,   \[1-\alpha = F_{\mathbf{u}}(\mathbf{x}),\]
(see Definition \ref{def:dirQ} and Equation \eqref{eq:ortProp})

Furthermore, Proposition \ref{prop:rotRegVar} implies that $F_{\mathbf{u}}$ belongs to the max-domain of attraction of a non-degenerate multivariate extreme value distribution $G_{\textbf{u}}$ with the same Frech\'et marginals (the latter given the tail equivalence among marginal distributions see \cite{resnick1}[Chapter 5]). Moreover, there exist two sequences $\mathbf{a}_{\mathbf{u}}(t)$, $\mathbf{b}_{\mathbf{u}}(t)$ such that,
\begin{equation}\label{eq:equivMaxConv}
\lim_{t\rightarrow\infty} t\left(1-F_{\mathbf{u}}\left(a_{\mathbf{u},j}(t)\, x_{j}+b_{\mathbf{u},j}(t);\ \ j=1,\ldots,d\right)\right)=-\ln\left(G_{\mathbf{u}}(\mathbf{x})\right).
\end{equation}
In addition, a direct consequence of \eqref{eq:equivMaxConv} is that each marginal of $G_{\mathbf{u}}$ has the form $\exp(-(1+\gamma x_{j})^{-1/\gamma}), \ \ j=1\ldots, d$, for $\gamma > 0$ (see also \cite{deHaan-ferreira}[Chapter 6]). Hence, it is possible to write,
\begin{equation}\label{eq:margConv}
\lim_{t\rightarrow\infty} t\left(1-F_{\mathbf{u},j}\left(a_{\mathbf{u},j}(t)x_{j}+b_{\mathbf{u},j}(t)\right)\right)=\left(1+\gamma x_{j}\right)^{-1/\gamma},
\end{equation}
where $F_{\mathbf{u},j}$ is the $j-$marginal of $F_{\mathbf{u}}$.
Thus, \eqref{eq:margConv} implies that for small values of  $\alpha$, the $(1-\alpha)-$quantile related to $F_{\mathbf{u},j}$ verifies the following relationship,
\begin{equation}\label{eq:univMathEst}
x_{\mathbf{u},j}(\alpha) \approx a_{\mathbf{u},j}(t)\frac{(1/t\alpha)^{\gamma}-1}{\gamma} + b_{\mathbf{u},j}(t),\quad \text{for all } j=1,\ldots,d.
\end{equation}

Now, to obtain the joint behavior characterizing the elements belonging to $\mathcal{Q}_{R_{\mathbf{u}}\mathbf{X}}(\alpha,\mathbf{e})$, we introduce the bivariate heuristic ideas in \cite{dehaan1995} based on a parameterizable scalar function $\rho_{\mathbf{u}}(\cdot)$ that approximates such joint quantile structure by a deformation of the marginal quantiles. Then, we first recall that any point $\mathbf{x}\in\mathbb{R}^{d}$ can be written alternatively in polar coordinates as $\mathbf{x}=||\mathbf{x}||\, (\mathbf{x}/||\mathbf{x}||)=\rho(\boldsymbol\theta) \, \boldsymbol\theta$, where $\rho(\boldsymbol\theta)\in \mathbb{R}^{+}$ and $\boldsymbol\theta$ belonging to the unit $d-$dimensional ball (for a further discussion see \cite{driver}[pg. 217]).  Note also that Assumption \textbf{\ref{assum2}} expressed in terms of the polar parameterization becomes in the analysis of upper-end points when $\boldsymbol\theta = (\theta_{1},\ldots,\theta_{d})$ is such that $0\leq\theta_{i}\leq 1$, for all $i=1,\ldots,d$.

Now, any point of $\mathcal{Q}_{R_{\mathbf{u}}\mathbf{X}}(\alpha,\mathbf{e})$ under polar parameterization will be denoted by $\mathbf{x}_{\mathbf{u}}(\alpha,\boldsymbol\theta)$, where $\boldsymbol\theta \in \Theta = \{(\theta_{1},\ldots,\theta_{d}) \text{ such that } ||\boldsymbol\theta|| = 1 \text{ and } 0\leq\theta_{i}\leq 1, \text{ for all } i=1,\ldots, d\}$. Finally,
we can write the following heuristic for the elements of $\mathcal{Q}_{R_{\mathbf{u}}\mathbf{X}}(\alpha,\mathbf{e})$,
\begin{equation}\label{eq:xTeo}
x_{\mathbf{u},j}(\alpha,\boldsymbol\theta) = a_{\mathbf{u},j}(t)\frac{(\rho_{\mathbf{u}}(\boldsymbol\theta)\theta_{j}/t\alpha)^{\gamma}-1}{\gamma} + b_{\mathbf{u},j}(t),\quad \text{for all } j=1,\ldots,d.
\end{equation}

It is crucial to remark the difference between $x_{\mathbf{u},j}(\alpha)$ given in \eqref{eq:univMathEst} and $x_{\mathbf{u},j}(\alpha,\boldsymbol\theta)$ in \eqref{eq:xTeo}. The former is related to the univariate quantile of the marginal $(R_{\mathbf{u}}\mathbf{X})_{j}$ and the latter is the $j-$component of an element in $\mathcal{Q}_{R_{\mathbf{u}}\mathbf{X}}(\alpha,\mathbf{e})$. Therefore, except for $\rho_{\mathbf{u}}(\boldsymbol\theta)$, all the elements in \eqref{eq:xTeo} are known or can be estimated. Then, the problem of estimating $x_{\mathbf{u},j}(\alpha,\boldsymbol\theta)$ turns into the problem of finding a mathematical expression for the scalar function $\rho_{\mathbf{u}}(\boldsymbol\theta)$. From \eqref{eq:equivMaxConv} and \eqref{eq:xTeo}, we obtain
\begin{equation}\label{eq:xDeduction}
\begin{aligned}
\alpha &= 1-F_{\mathbf{u}}(\mathbf{x}_{\mathbf{u}}(\alpha,\boldsymbol\theta)) \approx t^{-1}\left\{-\ln\left(G_{\mathbf{u}}\left(\frac{x_{\mathbf{u},j}(\alpha,\boldsymbol\theta)-b_{\mathbf{u},j}(t)}{a_{\mathbf{u},j}(t)};\ \ j=1,\ldots,d\right)\right)\right\}\\
& = t^{-1}\left\{-\ln\left(G_{\mathbf{u}}\left(\frac{(\rho_{\mathbf{u}}(\boldsymbol\theta)\theta_{j}/t\alpha)^{\gamma}-1}{\gamma};\ \ j=1,\ldots,d\right)\right)\right\}\\
& = -\frac{\alpha}{\rho_{\mathbf{u}}(\boldsymbol\theta)}\ln\left(G_{\mathbf{u}}\left(\frac{\theta_{j}^{\gamma}-1}{\gamma};\ \ j=1,\ldots,d\right)\right).
\end{aligned}
\end{equation}
Last equality in \eqref{eq:xDeduction} is due to the homogeneity property of $G_{\mathbf{u}}$ (see \cite{deHaan-ferreira} [Theorem 6.1.9]).

Hence, from \eqref{eq:xDeduction}, we achieve a expression of $\rho_{\mathbf{u}}(\boldsymbol\theta)$ by approximation. This expression will be denoted as
\begin{equation}\label{eq:rho}
\tilde{\rho}_{\mathbf{u}}(\boldsymbol\theta):=-\ln\left(G_{\mathbf{u}}\left(\frac{\theta_{j}^{\gamma}-1}{\gamma};\ \ j=1,\ldots,d\right)\right),
\end{equation}
which implies the approximation of $x_{\mathbf{u},j}(\alpha,\boldsymbol\theta)$ given by
\begin{equation}\label{eq:xApprox}
\tilde{x}_{\mathbf{u},j}(\alpha,\boldsymbol\theta) := a_{\mathbf{u},j}(t)\frac{(\tilde{\rho}_{\mathbf{u}}(\boldsymbol\theta)\,\theta_{j}/t\alpha)^{\gamma}-1}{\gamma} + b_{\mathbf{u},j}(t),\quad \text{for all } j=1,\ldots,d.
\end{equation}

Thus, $\mathcal{Q}_{\mathbf{X}}(\alpha,\mathbf{u})$ is approximated at high levels by the parameterization,
\begin{equation}\label{eq:ortPropAprox}
\tilde{\mathcal{Q}}_{\mathbf{X}}(\alpha,\mathbf{u},\boldsymbol\theta) := R'_{\mathbf{u}}\tilde{\mathcal{Q}}_{R_{\mathbf{u}}\mathbf{X}}(\alpha,\mathbf{e},\boldsymbol\theta),
\end{equation}
where $\tilde{\mathcal{Q}}_{R_{\mathbf{u}}\mathbf{X}}(\alpha,\mathbf{e},\boldsymbol\theta):=\left\{\tilde{\mathbf{x}}_{\mathbf{u}}(\alpha,\boldsymbol\theta),\, \boldsymbol\theta\in \Theta\right\}$.
%
Thus, by using previous characterizations one can get an \textit{out-sample} estimator for $\mathcal{Q}_{\mathbf{X}}(\alpha,\mathbf{u})$, which is the objective of next section.

\begin{Remark}\label{remarkAssum}
Note that the analysis provided in this section by using Assumptions \textbf{\ref{assum2}-\ref{assum4}} makes the characterization feasible independently of the chosen direction $\mathbf{u}$. However, if \textbf{\ref{assum3}} does not hold, but for some direction $\mathbf{u}$, $R_{\mathbf{u}}\mathbf{X}$ belongs to a multivariate max-domain of attraction, then the results of \cite{dehaan1995} for a general dimension $d$, can be directly applied for the transformed vector.

Conversely, our aim here is to develop a general directional theory where under the considered  Assumptions \textbf{\ref{assum2}-\ref{assum4}},  practitioners have: (1) to test the regular variation only once for the original random vector $\mathbf{X}$ and (2) the capacity to look at the data with different perspectives of analysis or even to apply an overall analysis, similar to those provided by density function or depth function approaches, by moving $\mathbf{u}$ in all its domain.
\end{Remark}

\section{Inference for \textit{DMQ} at high-levels}\label{sec:inference}

For any direction $\mathbf{u}$, let $R_{\mathbf{u}}\mathbf{X}_{1}, \dots, R_{\mathbf{u}}\mathbf{X}_{n}$ be independent and identically distributed  (\textit{i.i.d.}) random vectors,  distributed as $R_{\mathbf{u}}\mathbf{X}$ and denote by $\left\{[(R_{\mathbf{u}}\mathbf{X})_{j}]_{i:n}\right\}_{j=1}^{d},\, i=1,\ldots,n,$ the collection of the corresponding $n-$th order statistics for each marginal.

Marginal order statistics are important since they allow Equation  \eqref{eq:equivMaxConv}  to be written in terms of a subsample that provides significant information about the tail behavior of the joint distribution $F_{\mathbf{u}}$  (see \cite{deHaan-ferreira}[Section 7.2]). This subsample is related to a tuning intermediate sequence  $k:=k(n)\rightarrow\infty$, $k(n)/n\rightarrow 0$ as $n\rightarrow\infty$. This crucial sequence leads to the break point from which the information of an \textit{ordered sample} starts to be considered in the tail of the distribution. Then, we obtain,
\begin{equation*}\label{eq:estEquivMaxConv}
\lim_{n\rightarrow\infty} \frac{n}{k}\left(1-F_{\mathbf{u}}\left(a_{\mathbf{u},j}(n/k)\, x_{j}+b_{\mathbf{u},j}(n/k);\ \ j=1,\ldots,d\right)\right)=-\ln\left(G_{\mathbf{u}}(\mathbf{x})\right).
\end{equation*}
We now introduce an estimator of $\mathcal{Q}_{\mathbf{X}}(\alpha,\mathbf{u})$ (see Section \ref{subsec:qEstimator}) and we provide its asymptotic normality (see Section \ref{subsec:normality}). Furthermore,  we describe a bootstrap based methodology to find an optimal solution of the tuning parameter $k$ in our multivariate approach  (see Section \ref{subsec:bootstrap}).

Notice that due to Proposition \ref{prop:rotRegVar}, Proposition \ref{prop:rotRegVar2} and the quasi-orthogonal property in Equation \eqref{eq:ortProp}, all the results in the present section are fulfilled independently from the choice of $\mathbf{u}$. However, as we mentioned in Remark \ref{remarkAssum}, if we allow to lose the generality of the directional approach and we fix the direction $\mathbf{u}$ beforehand. Then, we could ask for the random vector $R_{\mathbf{u}}\mathbf{X}$ to belong to a multivariate max-domain of attraction and we fall on the extension of the work in \cite{dehaan1995} removing the tail equivalence for marginal distributions.

\subsection{Directional multivariate quantile estimator $\hat{\mathcal{Q}}_{\mathbf{X}}(\alpha,\mathbf{u})$}\label{subsec:qEstimator}
Note that if one has estimators for all the elements in \eqref{eq:xApprox}, an estimator of \eqref{eq:ortPropAprox} can be provided. For the estimation of the tail index $\gamma$, one can consider a $2-$steps procedure. Firstly, the multivariate regular variation condition in Assumption \textbf{\ref{assum3}} can be tested using the procedure described in \cite{einmahl-krajina}. This ensure the statistical equality of all the marginal tail indexes.

Secondly, each marginal estimation can be done through the moments estimators given in \cite{dekkers}, i.e.,
\begin{equation}\label{eq:tailIndex}
\hat{\gamma}_{j} := \mathbf{M}^{(1)}_{k, j}+1-\frac{1}{2}\left\{1-\left(\mathbf{M}^{(1)}_{k, j}\right)^{2}/\mathbf{M}^{(2)}_{k, j}\right\}^{-1},
\end{equation}
where $\mathbf{M}^{(r)}_{k, j} := \frac{1}{k}\sum_{i=0}^{k-1}\left\{\ln([(R_{\mathbf{u}}\mathbf{X})_{j}]_{n-i:n})-\ln([(R_{\mathbf{u}}\mathbf{X})_{j}]_{n-k:n})\right\}^{r},\, r=1,2$.

Notice that each $\hat{\gamma}_{j}$ depends on the sample size $n$ and the tuning parameter $k$. 
In Section \ref{subsec:bootstrap}, we discuss how to find an optimal selection of $k$ based on a joint estimation of the marginal tail indexes.

The estimators for the components of the sequences $\mathbf{a}_{\mathbf{u}}(n/k)$, $\mathbf{b}_{\mathbf{u}}(n/k)$ can be defined as in \cite{dehaan1995} by
\begin{align}\label{eq:secA}
\hat{a}_{\mathbf{u},j}(n/k) &:= [(R_{\mathbf{u}}\mathbf{X})_{j}]_{n-k:n}\mathbf{M}^{(1)}_{k, j}\max(1,1-\hat{\gamma}_{j}),\\
\label{eq:secB}
\hat{b}_{\mathbf{u},j}(n/k) &:= [(R_{\mathbf{u}}\mathbf{X})_{j}]_{n-k:n}.
\end{align}

The estimator of the scalar function $\tilde{\rho}_{\mathbf{u}}$ can be defined by
\begin{equation}\label{eq:empLimitDist}%
\begin{aligned}
\hat{\rho}_{\mathbf{u}}(\boldsymbol\theta) & := -\ln\left(\hat{G}_{\mathbf{u}}\left(\frac{\theta_{j}^{\hat{\gamma}_{j}}-1}{\hat{\gamma}_{j}};\ \ j=1,\ldots,d\right)\right),\\  \,\, \mbox{ where }
-\ln\left(\hat{G}_{\mathbf{u}}(\mathbf{x})\right) &= \frac{1}{k}\sum_{i=1}^{n}\boldsymbol 1_{\left\{\bigcup\limits_{j = 1}^{d}\left[ \left(R_{\mathbf{u}}\mathbf{X} _{i}\right)_{j}\, >\, \hat{a}_{\mathbf{u},j}(n/k)x_{\mathbf{u},j}+\hat{b}_{\mathbf{u},j}(n/k)\right]\right\}}.
\end{aligned}
\end{equation}
Hence, by using \eqref{eq:xApprox} and Equations \eqref{eq:tailIndex}-\eqref{eq:empLimitDist} one can get an estimator for elements of $\tilde{\mathcal{Q}}_{R_{\mathbf{u}}\mathbf{X}}(\alpha,\mathbf{e},\boldsymbol\theta)$:
\begin{equation}\label{eq:xEst}
\hat{x}_{\mathbf{u},j}(\alpha,\boldsymbol\theta,n/k) := \hat{a}_{\mathbf{u},j}(n/k)\left\{\frac{\left(\frac{k\,\hat{\rho}_{\mathbf{u}}(\boldsymbol\theta)}{n\,\alpha}\theta_{j}\right)^{\hat{\gamma}_{j}}-1}{\hat{\gamma}_{j}}\right\}+\hat{b}_{\mathbf{u},j}(n/k), \text{ for all } j=1,\ldots,d.
\end{equation}
We denote this set of points as $\hat{\mathcal{Q}}_{R_{\mathbf{u}}\mathbf{X}}(\alpha,\mathbf{e},\boldsymbol\theta,n/k)$ and using the quasi-orthogonal property, we get,
\begin{equation}\label{eq:ortPropEst}
\hat{\mathcal{Q}}_{\mathbf{X}}(\alpha,\mathbf{u},\boldsymbol\theta,n/k) = R'_{\mathbf{u}}\hat{\mathcal{Q}}_{R_{\mathbf{u}}\mathbf{X}}(\alpha,\mathbf{e},\boldsymbol\theta,n/k).
\end{equation}


\subsection{Asymptotic normality of $\hat{\mathcal{Q}}_{\mathbf{X}}(\alpha,\mathbf{u},\boldsymbol\theta,n/k)$}\label{subsec:normality}

We prove in the following the point-wise asymptotic normality of $\hat{\mathcal{Q}}_{\mathbf{X}}(\alpha,\mathbf{u},\boldsymbol\theta,n/k)$ in \eqref{eq:ortPropEst}, when $n\,\alpha\rightarrow 0$, as $n\rightarrow\infty$.   Firstly, in our setting, one can get
\begin{equation*}\label{eq:rewrite1}
\begin{aligned}
\sqrt{k}\,\left(\frac{\hat{a}_{\mathbf{u},j}(n/k)}{a_{\mathbf{u},j}(n/k)}-1\right)&\stackrel{d}{\rightarrow} A_{\mathbf{u},j},\\
\sqrt{k}\,\left(\frac{\hat{b}_{\mathbf{u},j}(n/k)-b_{\mathbf{u},j}(n/k)}{a_{\mathbf{u},j}(n/k)}-1\right)&\stackrel{d}{\rightarrow} B_{\mathbf{u},j},\\
\sqrt{k}\,\left(\hat{\gamma}_{j}-\gamma\right)&\stackrel{d}{\rightarrow} \Gamma_{\mathbf{u},j},\\
\sqrt{k}\left(-\log \hat{G}_{\mathbf{u}}(\mathbf{x}) + \log G_{\mathbf{u}}(\mathbf{x})\right)\stackrel{d}{\rightarrow} V_{\mathbf{u}}(\mathbf{x})& := W_{\mathbf{u}}(\mathbf{x}) (\mathbf{B}_{\mathbf{u}} + \mathbf{x}\odot \mathbf{A}_{\mathbf{u}})' \ \triangledown(-\log G_{\mathbf{u}}(\mathbf{x})),
\end{aligned}
\end{equation*}
where $\gamma > 0$, $\stackrel{d}{\rightarrow}$ means convergence in distribution, $\odot$ means a component-wise product,  $\triangledown(\cdot)$ means the vector of partial derivatives of a function; $\mathbf{A}_{\mathbf{u}}=(A_{\mathbf{u},j}; j=1,\ldots,d)$, $\mathbf{B}_{\mathbf{u}}=(B_{\mathbf{u},j}; j=1,\ldots,d)$ and $\boldsymbol\Gamma_{\mathbf{u}} = (\Gamma_{\mathbf{u},j}; j=1,\ldots,d)$ are such that,

\begin{equation*}\label{eq:rewrite2}
\begin{aligned}
A_{\mathbf{u},j} =& \gamma W_{\mathbf{u},j}(0) + \frac{(1-\gamma)^{2}(1-2\gamma)}{(1-4\gamma)}\left(\frac{P_{\mathbf{u},j}}{1-\gamma} + \frac{Q_{\mathbf{u},j}}{2}\right)\\ 
&- \frac{6(1-2\gamma)^{3}+2(1-\gamma)(1-2\gamma)^{2}-8(1-\gamma)^{3}}{2(1-4\gamma)^{2}(1-\gamma)(1-2\gamma)}\Gamma_{\mathbf{u},j},\\
B_{\mathbf{u},j} =& W_{\mathbf{u},j}(0),\\
\Gamma_{\mathbf{u},j} =& 2(1-\gamma)^{2}(1-2\gamma)P_{\mathbf{u},j} + \frac{(1-\gamma)^{2}(1-2\gamma)^{2}}{2}Q_{\mathbf{u},j},
\end{aligned}
\end{equation*}
where
\begin{equation*}\label{eq:rewrite3}
P_{\mathbf{u},j} = \int_{1}^{\infty} W_{\mathbf{u},j}(\mathbf{s})\frac{ds}{s} - W_{\mathbf{u},j}(\mathbf{0}), \quad
Q_{\mathbf{u},j} = 2\int_{1}^{\infty} W_{\mathbf{u},j}(\mathbf{s})\log(s)\frac{ds}{s} - 2W_{\mathbf{u},j}(\mathbf{0}),
\end{equation*}
$W_{\mathbf{u}}(\mathbf{x})$ is a zero-mean Gaussian random field with covariance function (see \cite{dhyr}),
\[Cov(W_{\mathbf{u}}(\mathbf{z}), W_{\mathbf{u}}(\mathbf{s})) = \mu_{\mathbf{u}}((\mathbf{0}, \mathbf{z}]^{c}\cap (\mathbf{0}, \mathbf{s}]^{c}).\]
and $\mu_{\mathbf{u}}$ is the $\sigma-$finite measure provided by Proposition \ref{prop:rotRegVar}, such that, $\mu_{\mathbf{u}}\left((\mathbf{0}, \mathbf{x}]^{c}\right) = -\log G_{\mathbf{u}}(\mathbf{x})$. Moreover, from Proposition 4.1 in \cite{dhyr}, one can see that the joint distribution of $(\mathbf{A}_{\mathbf{u}},\mathbf{B}_{\mathbf{u}},\boldsymbol\Gamma_{\mathbf{u}},W_{\mathbf{u}})$ is a multivariate Gaussian distribution. 
Thus, we can prove the following central limit theorem.

\begin{Proposition}[Point-wise asymptotic normality for $\hat{\mathcal{Q}}_{R_{\mathbf{u}}\mathbf{X}}$]\label{prop:normality}
Let $s_{n} := k/(n\,\alpha)$. Suppose that $\mathbf{X}$ satisfies  Assumption \textbf{\ref{assum4}} and  for $j=1, \ldots, d$ the following strong marginal second order condition holds
\[\lim_{n\rightarrow\infty} \frac{\frac{n}{k}\left[1-F_{\mathbf{u}}\left(\infty, \ldots, a_{\mathbf{u},j}(n/k)\frac{s_{n}^{\gamma}x_{j}-1}{\gamma} + b_{\mathbf{u},j}(n/k),\ldots, \infty\right)\right] + \frac{x_{j}}{s_{n}}}{\Lambda\left(\phi(n/k)\right)\psi_{\mathbf{u}}\left([-\infty, \infty]\times\cdots\times \left[\frac{s_{n}^{\gamma}x_{j}-1}{\gamma}, \infty\right]\times\cdots\times [-\infty, \infty]\right)} = 1.\]
 Assume that
\[\lim_{n\rightarrow\infty}(\log s_{n})/\sqrt{k} = 0,\quad\quad \lim_{n\rightarrow\infty} s_{n}\sqrt{k} \cdot \Lambda\left(\phi(n/k)\right) \psi_{\mathbf{u}}\left(\left[-\boldsymbol\infty,\frac{s_{n}^{\gamma}-\boldsymbol 1}{\gamma}\right]^{c}\right) = 0,\]
where $\phi(\cdot)$, $\Lambda(\cdot)$, $\psi(\cdot)$ are as in Definition \ref{def:2ndOrderRegVarying} and $\psi_{\mathbf{u}}(\cdot) := \psi\circ R_{\mathbf{u}}'(\cdot)$.

Then, for $n \rightarrow \infty$,
\[\sqrt{k}\left(\frac{\hat{x}_{\mathbf{u},j}(\alpha, \boldsymbol\theta, n/k) - x_{\mathbf{u},j}(\alpha, \boldsymbol\theta)}{\hat{a}_{\mathbf{u},j}(n/k)\int_{1}^{s_{n}}t^{\hat{\gamma}_{j}-1}(\log t)dt};\quad  j=1,\ldots,d\right),\]
converges in distribution to
\[\left(\left(\rho_{\mathbf{u}}(\boldsymbol\theta)\theta_{j}\right)^{\gamma}\Gamma_{\mathbf{u},j},\ \ j=1,\cdots,d\right).\]
\end{Proposition}
Proof of Proposition \ref{prop:normality} is given in  Appendix \ref{sec:proofs}. Finally, in Corollary \ref{corollary:AsympNormality} below, the point-wise asymptotic normality of $\hat{\mathcal{Q}}_{\mathbf{X}}(\alpha,\mathbf{u},\boldsymbol\theta,n/k)$ is derived since the orthogonal transformations preserve the result in Proposition~\ref{prop:normality}.
\begin{Corollary}[Point-wise asymptotic normality for $\hat{\mathcal{Q}}_{\mathbf{X}}$]\label{corollary:AsympNormality}
\begin{sloppypar}
The point-wise asymptotic normality property of the estimator $\hat{\mathcal{Q}}_{R_{\mathbf{u}}\mathbf{X}}(\alpha,\mathbf{e},\boldsymbol\theta,n/k)$ is preserved under orthogonal transformations. Therefore the quasi-orthogonal property in \eqref{eq:ortPropEst} implies the point-wise asymptotic normality of $\hat{\mathcal{Q}}_{\mathbf{X}}(\alpha,\mathbf{u},\boldsymbol\theta,n/k)$.
\end{sloppypar}
\end{Corollary}

%
%

\subsection{Bootstrap method to estimate the tuning parameter $k=k(n)$}\label{subsec:bootstrap}
From Equations \eqref{eq:tailIndex}-\eqref{eq:empLimitDist}, one  can appreciate the key role of sequence $k=k(n)$. However, it is not an easy task to establish optimal tuning parameter $k$ for a given sample size $n$. This tuning parameter is complicated to tackle in practice and methods to provide optimality are still a matter of research and discussion.

In the recent literature, one can find only heuristic guidelines adapted to each multivariate application (e.g. \cite{cedh,cedhz,bernardino4}), where the selection of parameters such as $k$ are mostly based on the identification of a common region of stability across the estimation of particular marginal elements such as marginal tail indexes. For instance, $k$ can be selected through a graphical visualization of the common range of values providing the flattest behavior around the marginal estimations. On the other hand, some sophisticated methodologies based on bootstrap have been presented in the univariate case to provide optimality on the choice of $k$ (e.g. \cite{draisma,danielsson,ferreira,qi}). Therefore, a natural question here is: how to select an optimal value of $k$ to perform estimation in the multivariate context? To achieve this goal, it is necessary to overcome the lack of a total order in $\mathbb{R}^{d}$, for $d\geq 2$. In this sense, we will use the orthant order introduced in \cite{torres1st}, which is a partial order in $\mathbb{R}^{d}$ based on Definition \ref{def:orthant}. For a fixed direction $\mathbf{u}$,
\[\mathbf{x} \preceq_{\mathbf{u}} \mathbf{y},\quad\text{ if and only if, }\quad \mathfrak{C}_{\mathbf{x}}^{\mathbf{u}}\supseteq\mathfrak{C}_{\mathbf{y}}^{\mathbf{u}},\]
where $\mathbf{x},\mathbf{y}\in \mathbb{R}^{d}$ and $\mathfrak{C}_{\mathbf{x}}^{\mathbf{u}}$ is as in Definition \ref{def:orthantQR}. Equivalently,
\begin{equation}\label{eq:porder}
\mathbf{x} \preceq_{\mathbf{u}} \mathbf{y},\quad\text{ if and only if,}\quad R_{\mathbf{u}}\mathbf{x}\leq R_{\mathbf{u}}\mathbf{y},
\end{equation}
where the inequality on the right side is component-wise. Our proposal is described in the following pseudo-algorithm, which is based on the univariate method introduced in \cite{danielsson}.
\begin{enumerate}[Step 1.]
\item Rotate the sample to generate $\{R_{\mathbf{u}}\mathbf{x}_{1},\ldots,R_{\mathbf{u}}\mathbf{x}_{n}\}$.

\item Set $m_{1} = \lfloor n^{1-\epsilon}\rfloor$ for some $\epsilon \in (0, 1/2)$, where $\lfloor\cdot\rfloor$ denotes the integer part function. Draw a large number $B_{1}$ of bootstrap samples of size $m_{1}$ and drop the observations with non-positive components (this is equivalent to keep observations greater than $\mathbf{0}$ according to \eqref{eq:porder}). After this, use the marginal order to sort each of the remaining observations of the bootstrap samples.

\item Denote by $Err_{j}(m_{1}, b_{1}, k_{j})$ the error obtained in each marginal $j = 1,\ldots,d$, where $k_{j}$ varies from $1$ to $m_{1}-1$,
\begin{equation*}
Err_{j}(m_{1}, b_{1}, k_{j}) := \left(\mathbf{M}^{(2)}_{k_{j}, j} - 2\left(\mathbf{M}^{(1)}_{k_{j}, j}\right)^{2}\right)^{2}, \quad b_{1} = 1,\ldots,B_{1}.
\end{equation*}
Then, determine the value $k_{j}(m_{1})$ that minimizes
\[\frac{1}{B_{1}}\sum_{b_{1}=1}^{B_{1}} Err_{j}(m_{1}, b_{1}, k_{j}).\]

\item Set $m_{2} = \lfloor m_{1}^{2}/n\rfloor$, and repeat Step 2 and Step 3 to obtain $k_{j}(m_{2})$.

\item  Estimate marginal rates of convergence to the tail index $\gamma$ by
\[\hat{\pi}_{j} = \log\left(\frac{k_{j}(m_{1})}{-2 \log(m_{1}) + 2\log(k_{j}(m_{1}))}\right),\]
which marginally are consistent estimators (see \cite{qi}).

\item The optimal selection for $k = k(n)$ is given by,
\begin{equation*}\label{eq:kEst}
\hat{k}(n) := \frac{1}{d}\sum_{j = 1}^{d}\frac{k_{j}(m_{1})^{2}}{k_{j}(m_{2})}\left(1 - \frac{1}{\hat{\pi}_{j}}\right)^{1/(2\hat{\pi}_{j}-1)}.
\end{equation*}
\end{enumerate}
Remark 2 in \cite{qi} should be applied for $n < 2000/2^{d}$.

Notice that the previous bootstrap method  selects the optimal $k$ in terms of the quality of the approximation of the tail index $\gamma$. However it does not imply that the same  $k$ will be optimal in the estimation of $\tilde{\rho}_{\mathbf{u}}$ in Equation~\eqref{eq:empLimitDist}.

\section{Simulation Study}\label{sec:tExample}
In this section, we illustrate the estimation methodology introduced in Section \ref{sec:inference} by using  the  $d$-dimensional $t$-distribution with d.f. $\nu$. This distribution satisfies Assumptions \textbf{\ref{assum2}-\ref{assum4}} with  multivariate   regular variation indexes  $(\gamma,\, \pi)=(1/\nu,\, -2/\nu)$ (see \cite{hult}, \cite{hua}). To derive the theoretical \textit{DMQ} for any direction $\mathbf{u}$, it is necessary to recall Lemma 3.1 in \cite{hult} for  elliptical distributions (see Lemma \ref{lemma:elliptic}).

Indeed, Lemma \ref{lemma:elliptic} establishes that $R_{\mathbf{u}}\mathbf{X}$ is again a multivariate $t-$distribution with the same d.f. $\nu$, but with location and scale given by $\boldsymbol\mu_{\mathbf{u}} = R_{\mathbf{u}} \boldsymbol\mu$, $\Sigma_{\mathbf{u}} = R_{\mathbf{u}}\Sigma R_{\mathbf{u}}'$. Thus, tail indexes remain invariant for all $\mathbf{u}$ and it makes this model suitable to perform simulation analysis in any direction and/or in different dimensions. Some results in 2D and 3D scenarios are derived using the following $t-$distributions,
\begin{equation}\label{eq:distExample}
\begin{aligned}
\quad\boldsymbol\mu = [0,0]',\qquad &\qquad\Sigma = \begin{bmatrix}
5   & 0.1\\
0.1 &   1
\end{bmatrix},\qquad &\nu=3.\quad\\
\boldsymbol\mu = [0,0,0]',\qquad &\Sigma = \begin{bmatrix}
5 & 2.44 & -1.88\\
2.44 & 2.12 & 0.04\\
-1.88 & 0.04 & 2.36
\end{bmatrix},\qquad &\nu=4.
\end{aligned}
\end{equation}

The main goal of this section is to illustrate the differences and the importance of the directions in our methodology, as well as, to show the performance of the estimation method for $\mathcal{Q}_{\mathbf{X}}(\alpha,\mathbf{u})$. Therefore, the initial analysis over $\mathbf{X}$ is performed in the classical direction $\mathbf{e}$. Later,  the direction given by the main axis of the elliptical random vector is considered, which is equivalent to the vector characterizing the first principal component (\textit{FPC}). We focus the initial part of the study to the bivariate case and then we present the results for $d = 3$.

\begin{figure}[htbp]
\begin{center}
\includegraphics[height=6cm,width=6cm]{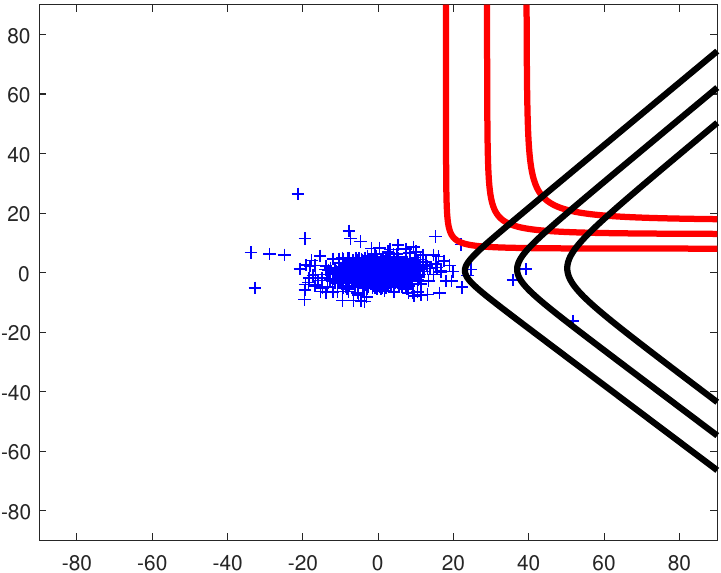}\vspace{-0.25cm}
\caption{Classical and \textit{FPC} directional quantiles in the bivariate $t-$distribution case for $\alpha = 1/500$, $1/2000$ and $1/5000$.}\label{fig:exampleT}
\end{center}
\end{figure}

Figure \ref{fig:exampleT} shows in red the theoretical curves $Q_{\mathbf{X}}(\alpha, \mathbf{e}) \equiv \{\mathbf{x}\, | \, F_{\mathbf{e}}(\mathbf{x})= 1-\alpha\}$ for three extreme values of $\alpha$ ($1/500,\, 1/2000,\, 1/5000$). It is also displayed in black the theoretical curves $Q_{\mathbf{X}}(\alpha, FPC) \equiv R_{FPC}'\{\mathbf{x}\, | \, F_{FPC}(\mathbf{x})= 1-\alpha\}$ for the same $\alpha$'s, (in this case the theoretical $FPC$ is $(0.9997,\, 0.025))$. These level curves show visual improvements of the extreme detection through directional analysis, since \textit{FPC} does take into account the shape of the data in contrast to the classical direction $\mathbf{e}$.

Now, we proceed to describe step by step all the necessary elements for the estimation of the extreme \textit{DMQ}. Our simulation study has been done for different sample sizes, but we present here only two representative cases: (1) $n = 500$ (``small sample''), and (2) $n = 5000$ (``large sample''). We consider in the following  $\alpha = 1/n$.

\begin{enumerate}
\item \textbf{Tuning parameter} $k(n)$: This parameter is estimated with the bootstrap methodology described in Section \ref{subsec:bootstrap} by considering $1000$ bootstrap samples. We performed $100$ iterations of the estimation procedure, i.e., $100$ samples from the $t-$model are drawn, then the optimal bootstrap selection of $k$ is done and corresponding $\hat{\gamma}_{j},\, j=1,2$ are calculated. Figure \ref{fig:kEst} displays box-plots  of the optimal $k$ when $n \in\{500,\, 5000\}$. 

\begin{figure}[htbp]
\begin{center}
\includegraphics[height=4.5cm,width=6cm]{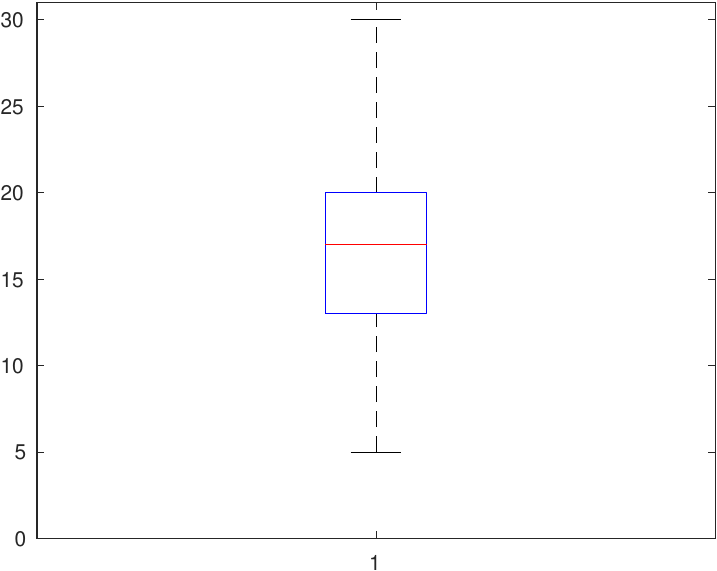}
\includegraphics[height=4.5cm,width=6cm]{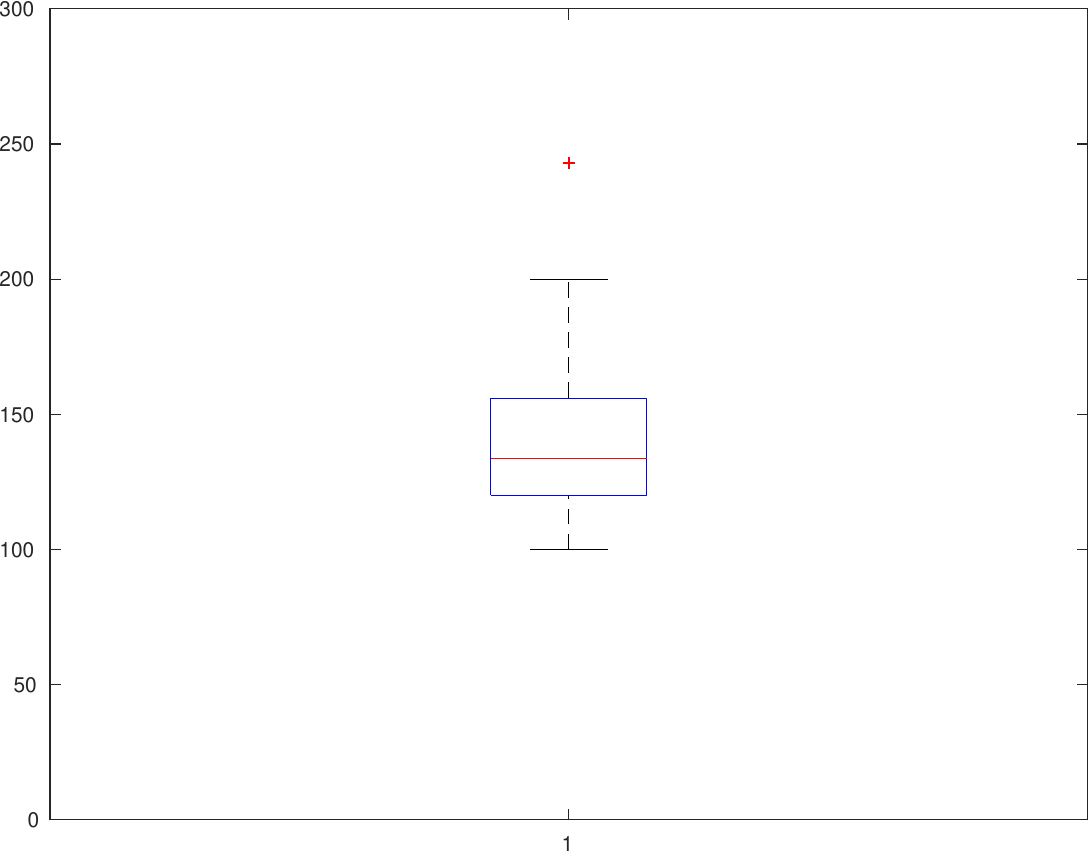}\vspace{-0.25cm}
\centerline{(A)\; $n = 500$ \hspace{5cm} (B)\; $n = 5000$}
\caption{Boxplots of the bootstrap estimation of the tuning parameter $k$.  }\label{fig:kEst}
\end{center}
\end{figure}

Similarly, Figure \ref{fig:tailIndex} shows the results obtained for the ratio of $\hat{\gamma}_{1}/\gamma$ when $n \in\{500,\, 5000\}$. The results for the other marginal are similar. However, since $\Sigma_{11} > \Sigma_{22}$ in \eqref{eq:distExample}, we only display here the ratios of the first marginal. 

\begin{figure}[ht!]
\begin{center}
\includegraphics[height=4.5cm,width=6cm]{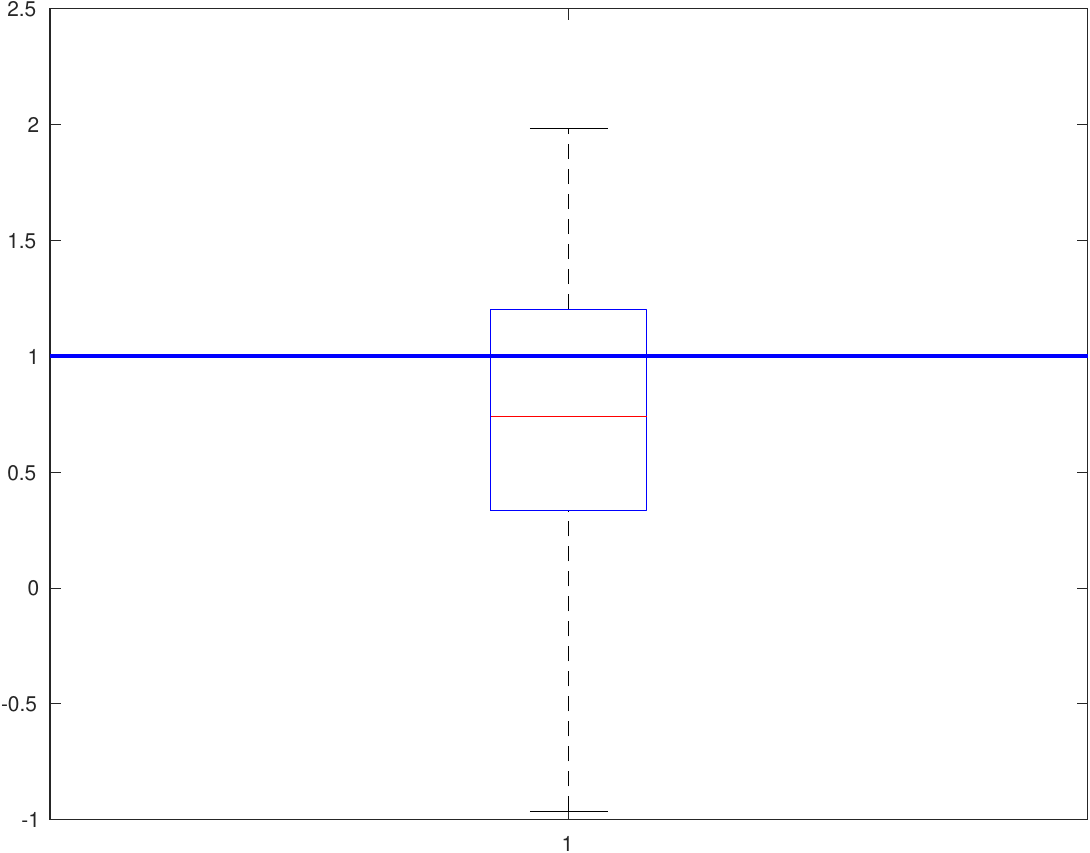}
\includegraphics[height=4.5cm,width=6cm]{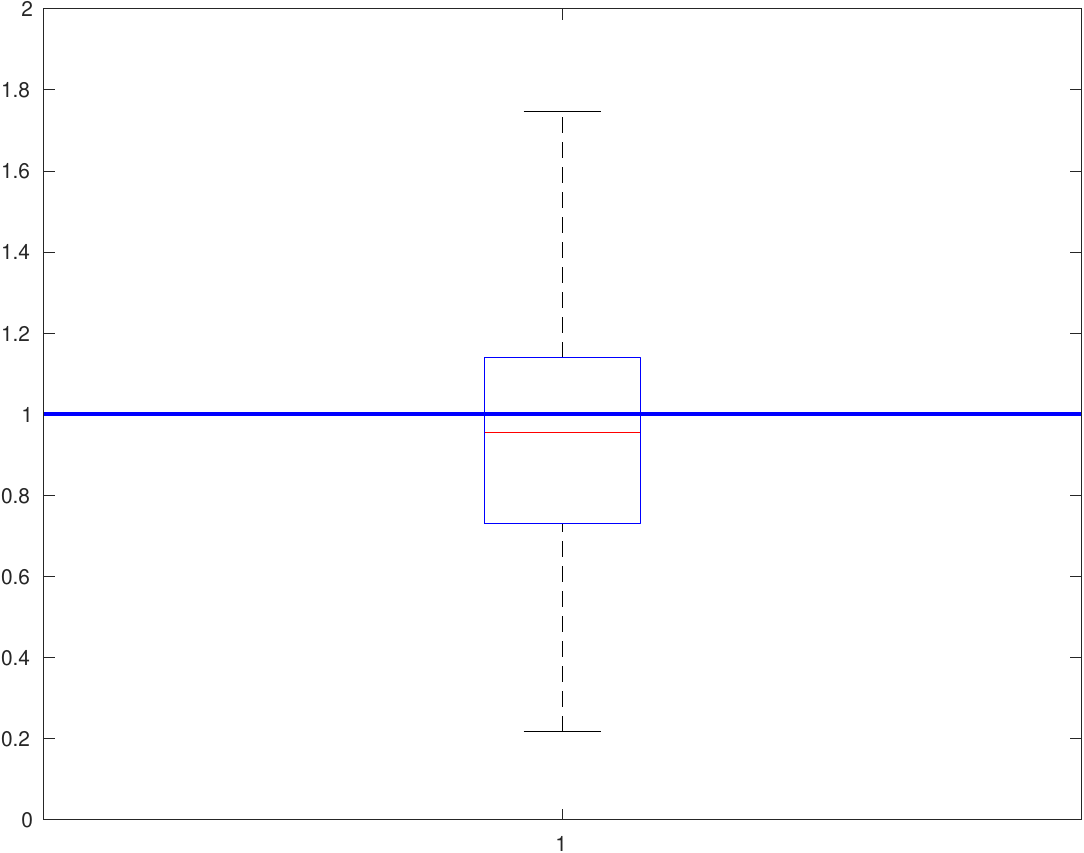}\vspace{-0.25cm}
\centerline{(A)\; $n = 500$ \hspace{5cm} (B)\; $n = 5000$}
\caption{Boxplots for the ratio $\hat{\gamma}_{1}/\gamma$.  }\label{fig:tailIndex}
\end{center}
\end{figure}

\item  \textbf{Tail index estimation} $\gamma$ \textbf{and sequences of normalization} $a_{\mathbf{u},j},\, b_{\mathbf{u},j}$: Previous step provides estimations of the marginal tail indexes $\hat{\gamma}_{j}$ through moments estimators in \eqref{eq:tailIndex} by using the optimal bootstrap selection of $k$ described in Section \ref{subsec:bootstrap}. Then, for this simulated data example and without loss of generality, we complete the estimation of $\gamma$ by taking the average of the marginal estimations. Finally, the estimations of the sequences of normalization $a_{\mathbf{u},j}$ and $b_{\mathbf{u},j}$ are given by using \eqref{eq:secA} and \eqref{eq:secB}.

\item \textbf{The scalar function} $\tilde{\rho}_{\mathbf{u}}(\boldsymbol\theta)$: Note that \eqref{eq:rho} uses the function $-\ln(G_{\mathbf{u}}(\cdot))$, which is the stable tail dependence function of the multivariate extreme value distribution $G_{\mathbf{u}}$. The theoretical tail function for a multivariate $t-$distribution is provided by  \cite{niko} [Theorem 2.3]. For sake of clarity this result is recalled in Appendix \ref{sec:proofs} (see Theorem \ref{remarkNiko}).

Therefore for any direction $\mathbf{u}$, we can calculate $\tilde{\rho}_{\mathbf{u}}(\boldsymbol\theta)$ and its estimator by using Lemma \ref{lemma:elliptic}, Theorem \ref{remarkNiko} and Equation \eqref{eq:empLimitDist}. Figure \ref{fig:rhoEst} shows the theoretical curves  $\tilde{\rho}_{\mathbf{e}}(\boldsymbol\theta)$ (in magenta) and the estimated ones (in blue), with the argument $\boldsymbol\theta$ described in terms of the $(d-1)$ angles of its polar parameterization. We can appreciate a good performance of the estimation for both sample sizes. Furthermore, as expected, the larger the sample size, the better the performance of the estimator.

\begin{figure}[ht!]
\begin{center}
\includegraphics[height=4.5cm,width=6cm]{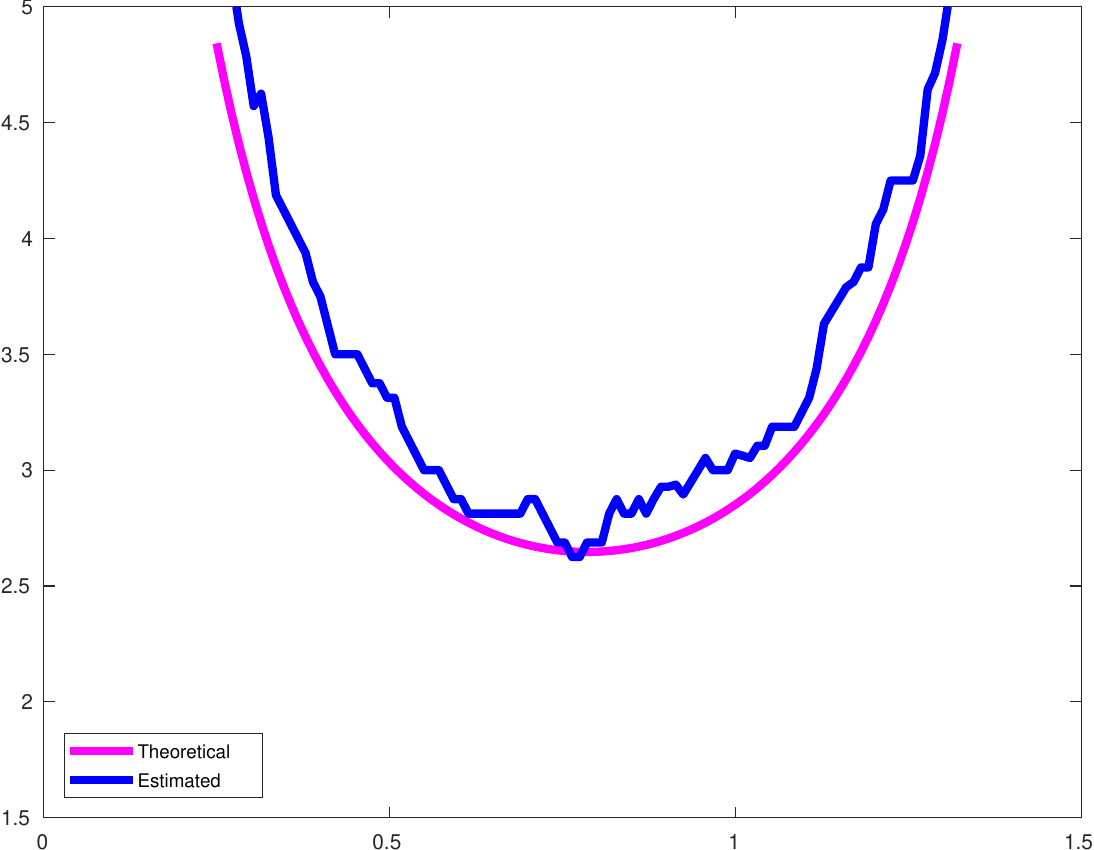}
\includegraphics[height=4.5cm,width=6cm]{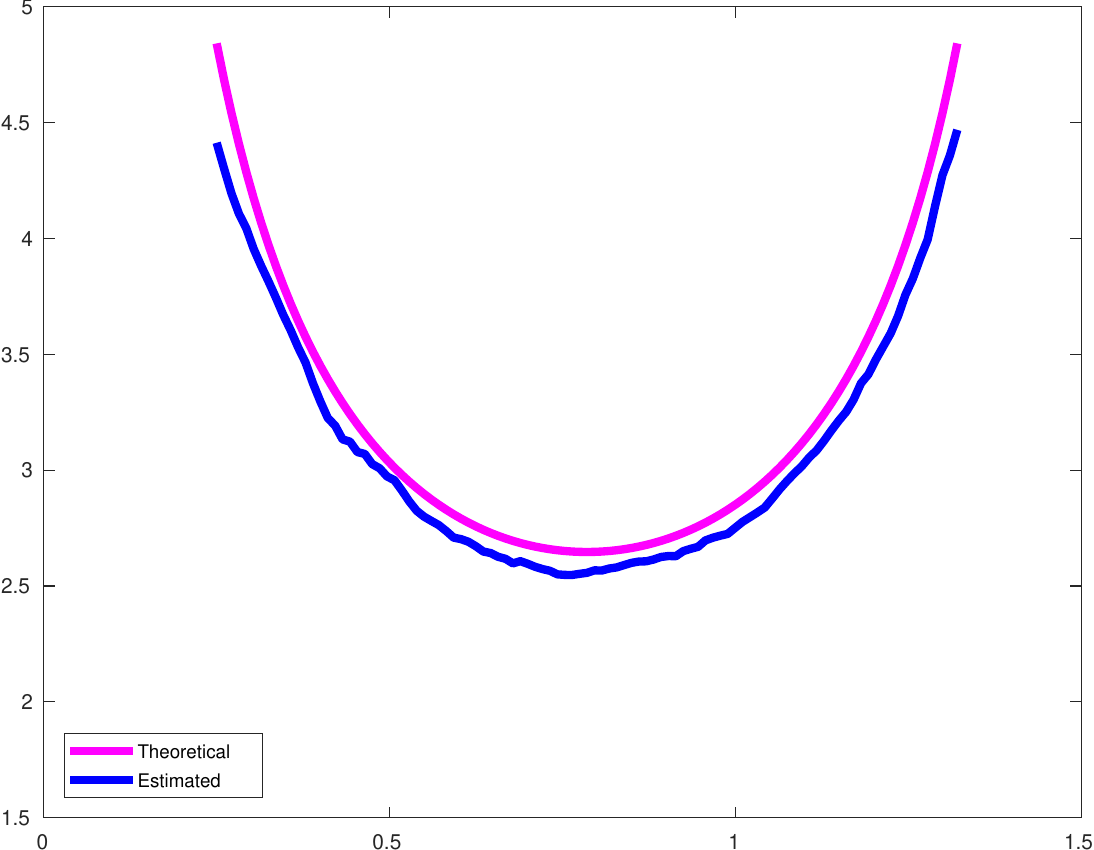}\vspace{-0.25cm}
\centerline{(A)\; $n = 500$ \hspace{5cm} (B)\; $n = 5000$}
\caption{Theoretical and estimated curves $\rho_{\mathbf{e}}(\boldsymbol\theta)$.}\label{fig:rhoEst}
\end{center}
\end{figure}

\item \textbf{The directional quantile curve} $\mathcal{Q}_{\mathbf{X}}(1/n,\mathbf{e})$: Once the previous steps are completed, both theoretical and estimated results for $\mathcal{Q}_{\mathbf{X}}(1/n,\mathbf{e})$ can be calculated. We use the $t-$model to simulate $100$ Monte Carlo samples and we apply previous  items 1-3 in order to construct point-wise confidence bands. The results are displayed in Figure \ref{fig:estQ}.
\begin{figure}[ht!]
\begin{center}
\includegraphics[height=6cm,width=7cm]{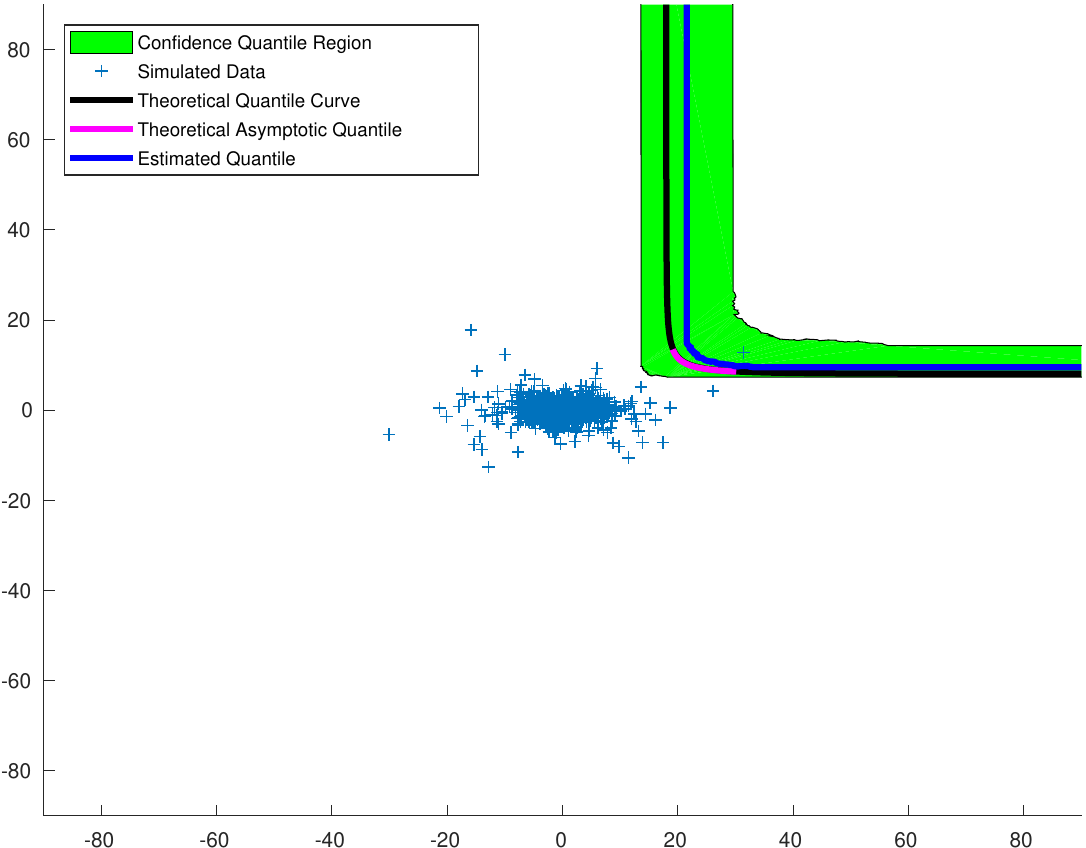}
\includegraphics[height=6cm,width=7cm]{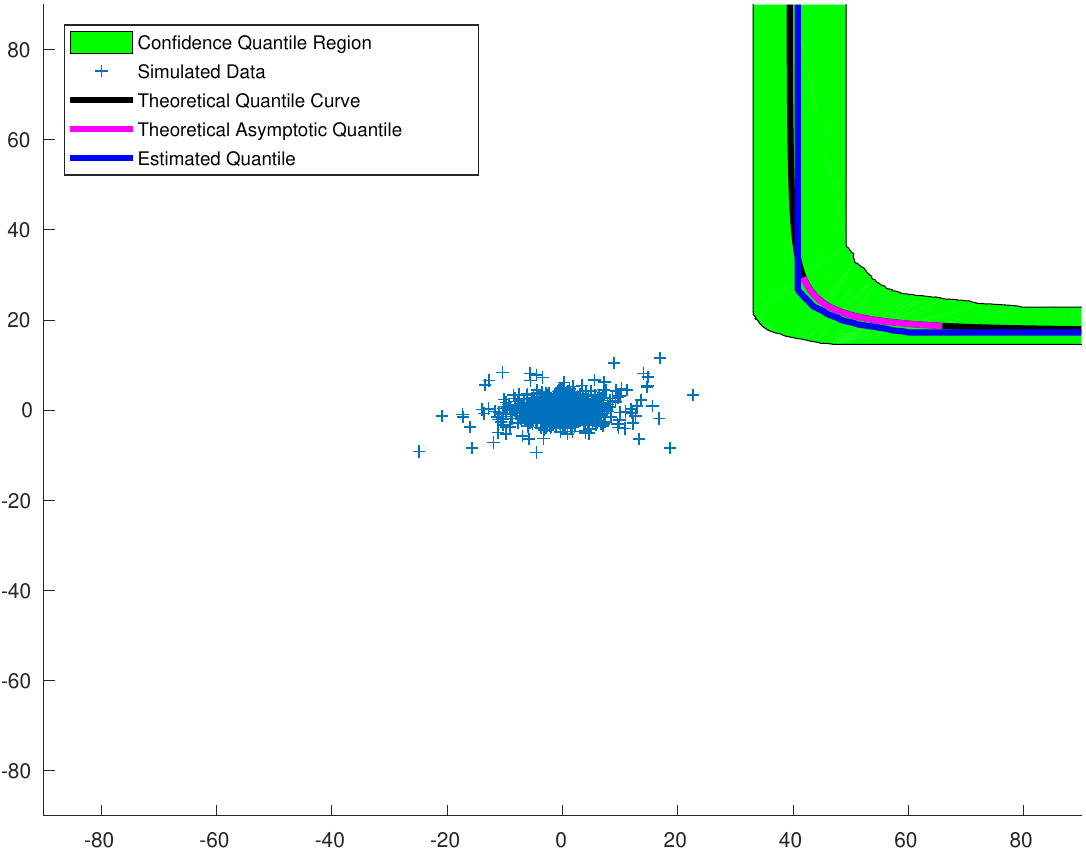}
\vspace{-0.25cm}
\centerline{(A)\;$n=500$ \hspace{5cm} (B)\;$n=5000$}
\caption{Estimations for $\mathcal{Q}_{\mathbf{X}}(1/n,\mathbf{e})$.}
\label{fig:estQ}
\end{center}
\end{figure}

Then, for both sample sizes $n\in\{500,\, 5000\}$, Figure \ref{fig:estQ} shows the theoretical quantiles plotted in black\footnote{Theoretical quantiles are the computational $1-\alpha$ iso-curves and iso-surfaces of the multivariate $t-$distribution.}, theoretical asymptotic  approximations through the tail function are in magenta, medians point by point of the 100 Monte Carlo estimated curves are in blue and confidence regions from $15\%$ to $85\%$ are shaded in green. We can appreciate the accuracy of the estimations of $\mathcal{Q}_{\mathbf{X}}(1/n,\mathbf{e})$.

Now, for the \textit{FPC} direction, we have that theoretical \textit{FPC} is equal to $(0.9997,\, 0.025)$ and the parameters of this directional model  are
\[\boldsymbol\mu_{FPC} = [0,0]',\qquad \Sigma_{FPC} = \begin{bmatrix}
3.0001  &  2.0025\\
2.0025  &  2.9999
\end{bmatrix}.\]
Thereby, we calculate theoretical $\mathcal{Q}_{R_{\mathbf{FPC}}\mathbf{X}}(1/n,\mathbf{e})$ and associated estimators by using previous items 1-4. We  consider  the same sample sizes $n \in\{500,\, 5000\}$  and Figure \ref{fig:estQRot} presents the results in the same colors as before.

\begin{figure}[ht!]
\begin{center}
\includegraphics[height=6cm,width=7cm]{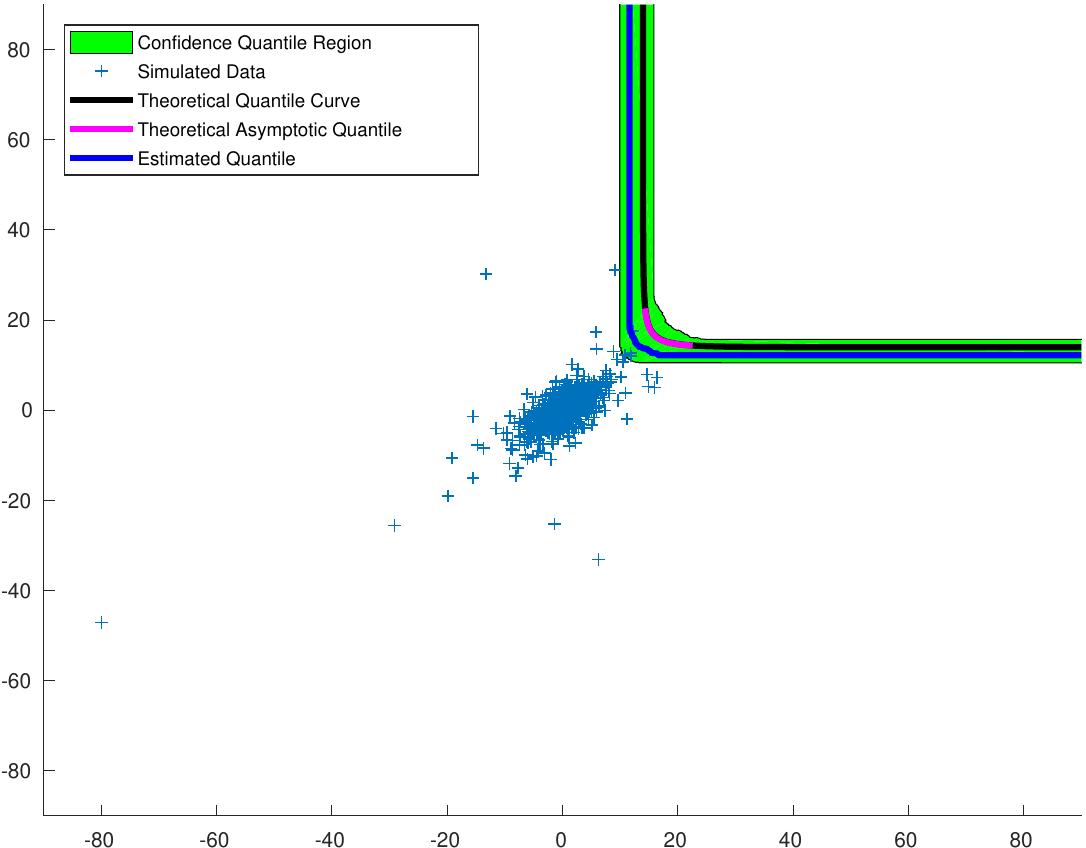}
\includegraphics[height=6cm,width=7cm]{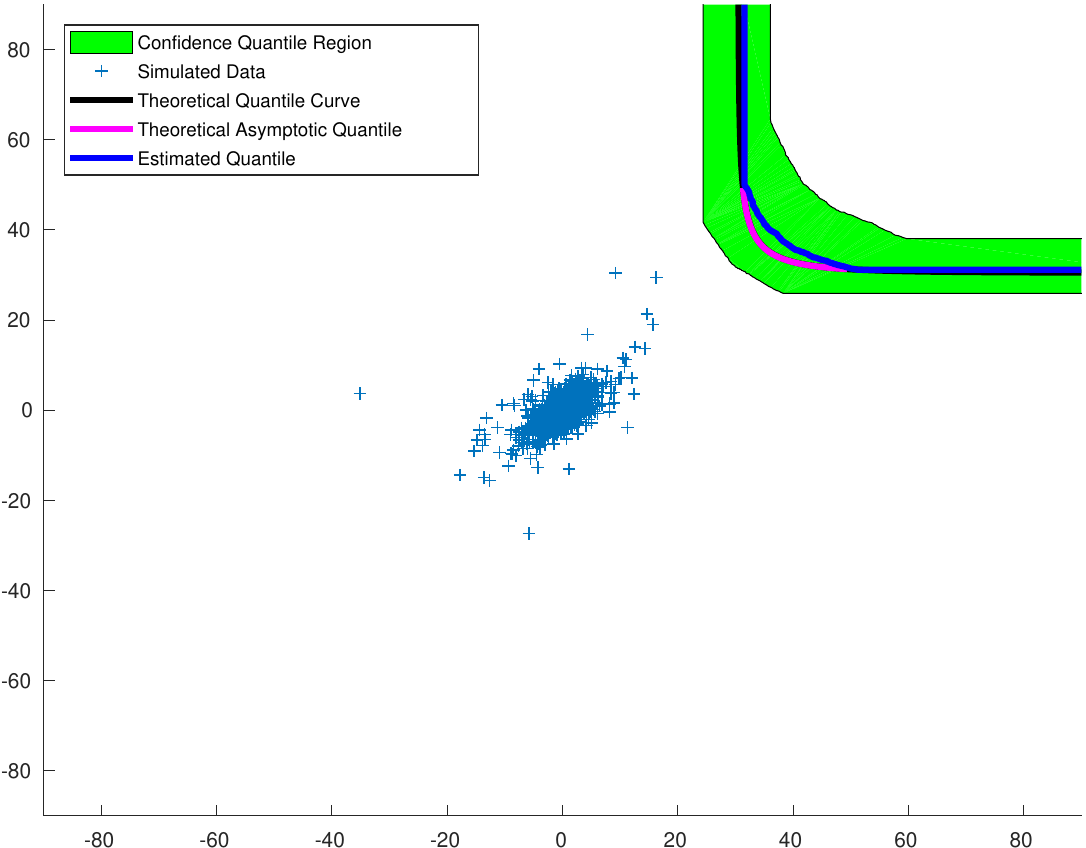}
\vspace{-0.25cm}
\centerline{(A)\;$n=500$\hspace{5cm} (B)\;$n=5000$}
\caption{Estimations for $\mathcal{Q}_{R_{\mathbf{u}}\mathbf{X}}(1/n,\mathbf{e})$ with $\mathbf{u} = FPC$.}
\label{fig:estQRot}
\end{center}
\end{figure}
\begin{sloppypar}
Finally, applying the inverse rotation indicated in Equation \eqref{eq:ortPropEst}, Figure \ref{fig:estQFin} presents the results for $\mathcal{Q}_{\mathbf{X}}(1/n,(0.9997,\, 0.025))$. One can appreciate a good performance of our estimators and the improvements on the visualization of extremes, which are more in concordance with the shape of this data-set.
\end{sloppypar}
\begin{figure}[ht!]
\begin{center}
\includegraphics[height=6cm,width=7cm]{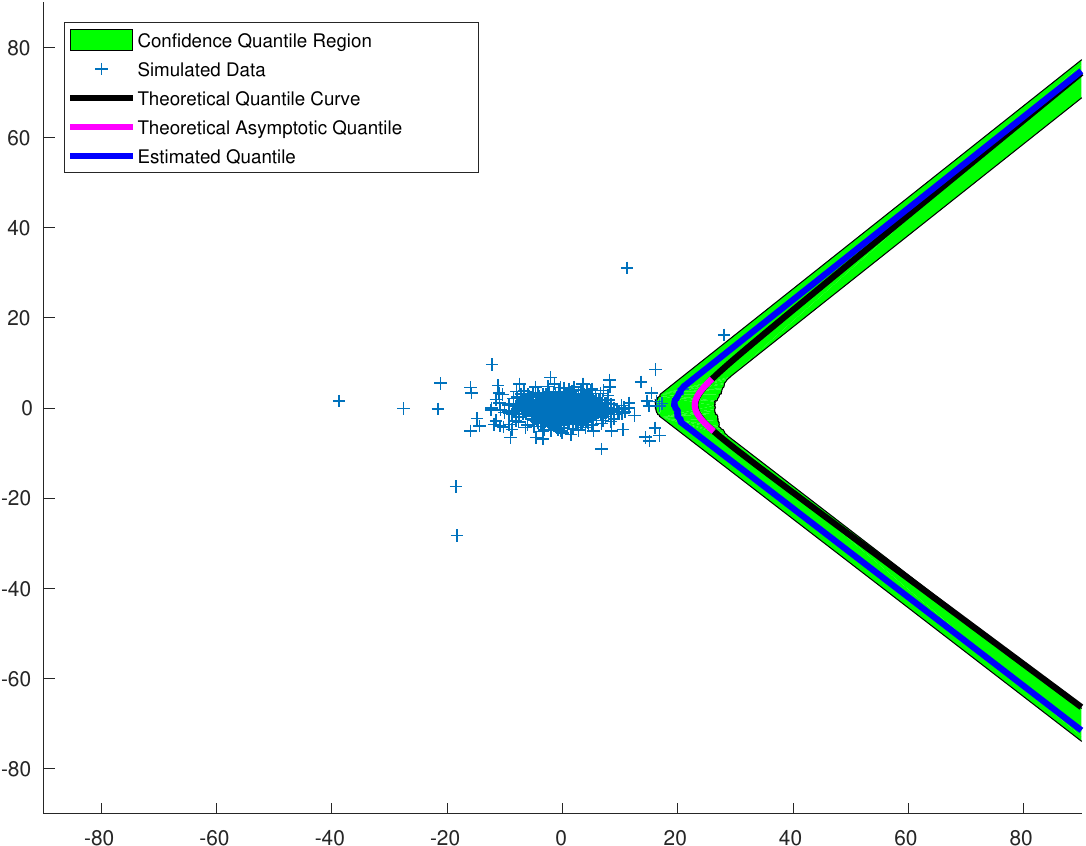}
\includegraphics[height=6cm,width=7cm]{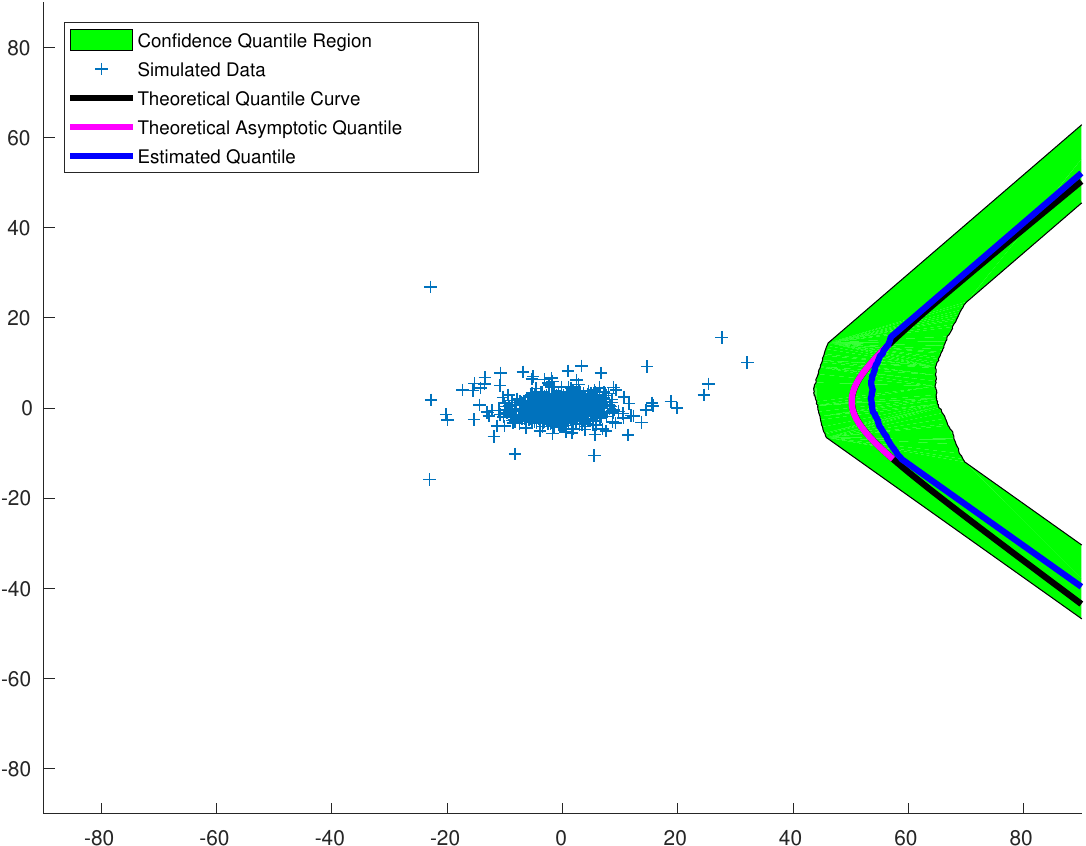}
\vspace{-0.25cm}
\centerline{(A)\;$n=500$\hspace{5cm} (B)\;$n=5000$}
\caption{Estimations for $\mathcal{Q}_{\mathbf{X}}(1/n,FPC)$.}\label{fig:estQFin}
\end{center}
\end{figure}
\end{enumerate}

\begin{figure}[ht!]
\begin{center}
\includegraphics[height=6cm,width=6cm]{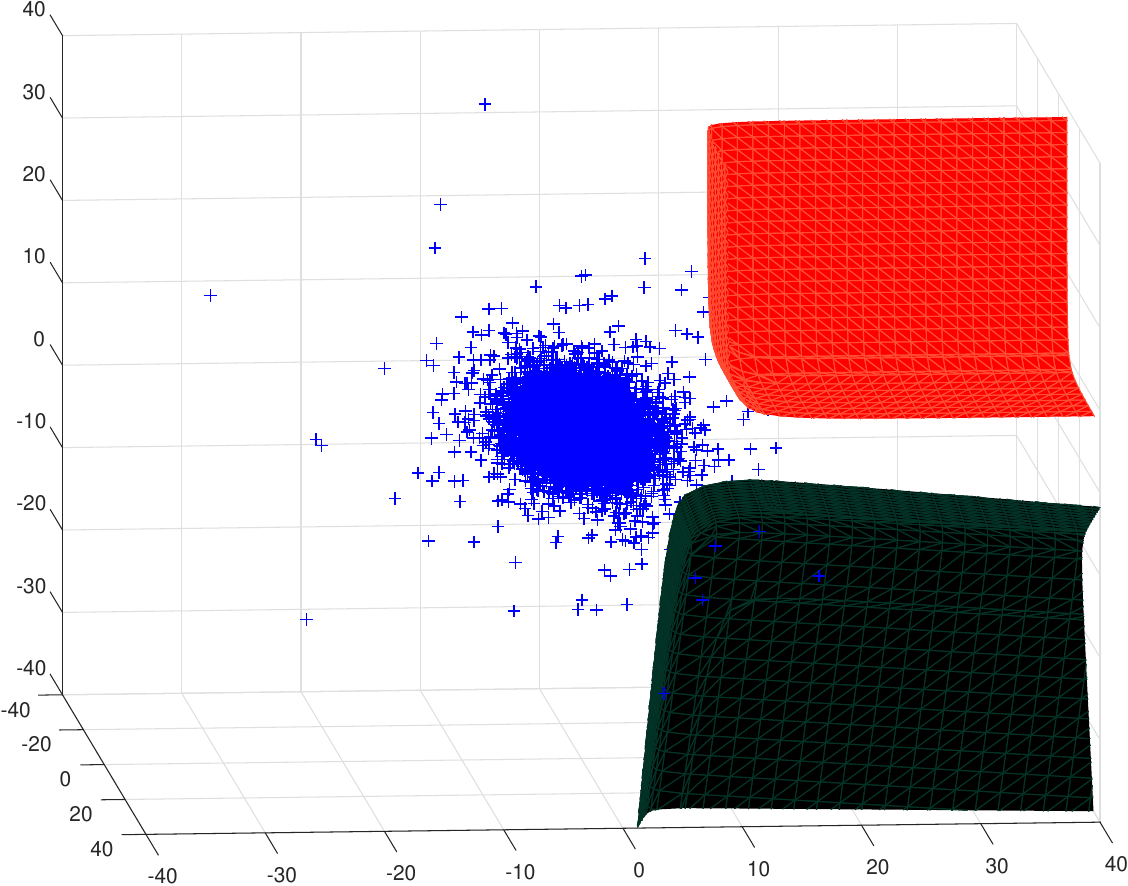}\vspace{-0.25cm}
\caption{Classical and \textit{FPC} directional quantiles for level  $\alpha = 1/500$.}\label{fig:qTeo3D}
\end{center}
\end{figure}

Similar results can be shown for the case $d=3$ presented in \eqref{eq:distExample}. Firstly, Figure \ref{fig:qTeo3D} displays the theoretical quantile surfaces at level $\alpha = 1/500$. The red quantile is performed in the classical direction $\mathbf{e}$ and the black one is in the $FPC$ direction. These iso-surfaces shown more concordance with the shape of the data when the $FPC$ direction is considered.

\begin{figure}[ht!]
\begin{center}
\includegraphics[height=4.5cm,width=6cm]{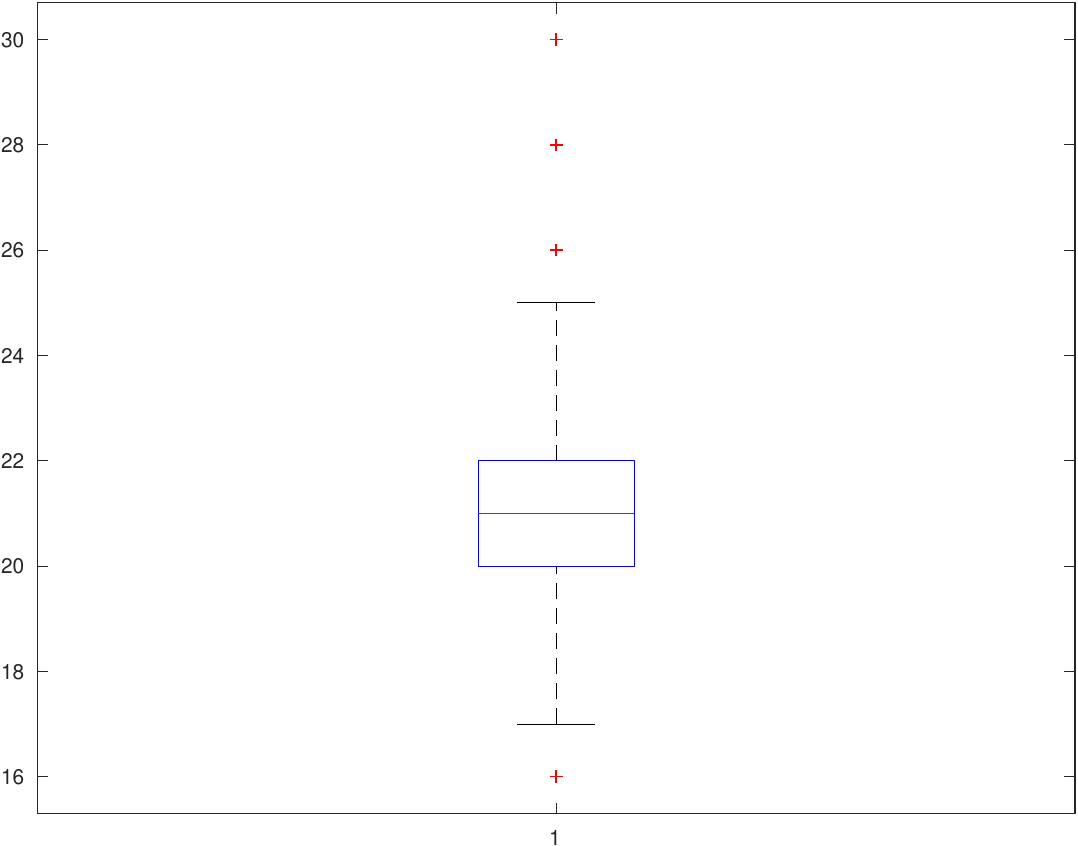}
\includegraphics[height=4.5cm,width=6cm]{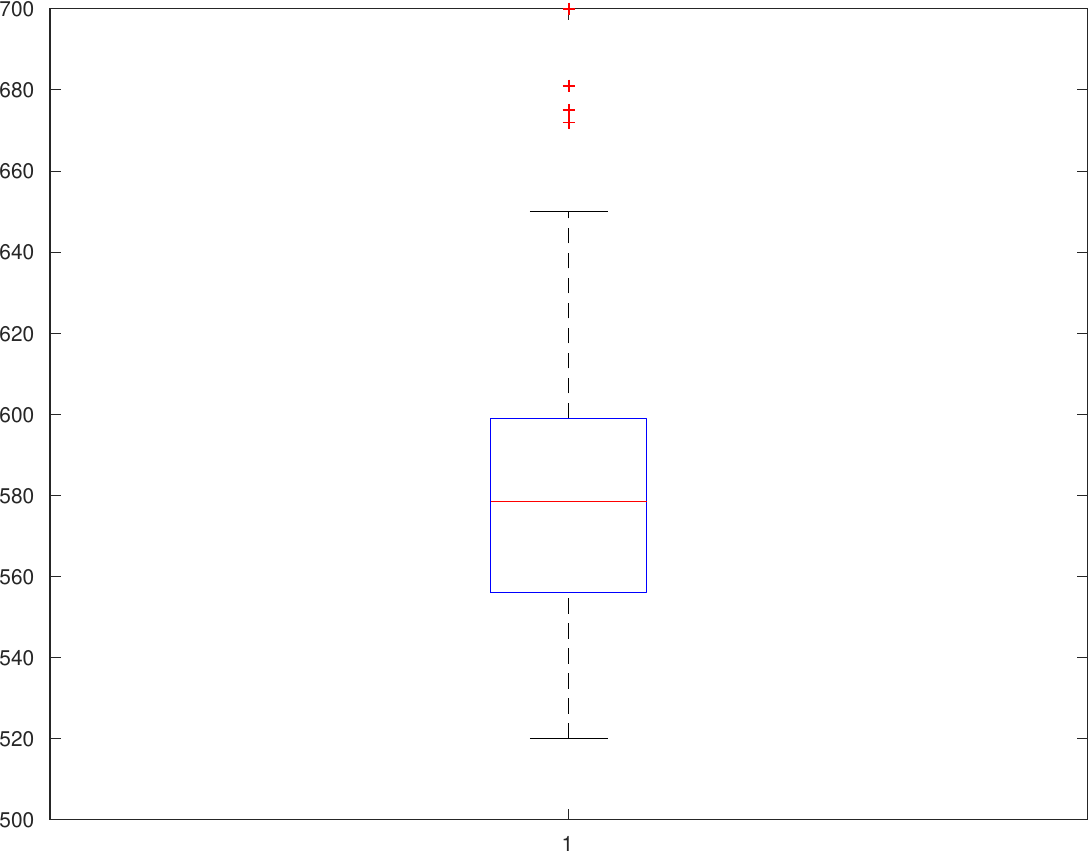}
\vspace{-0.25cm}
\centerline{(A)\; $n = 500$ \hspace{4cm} (B)\; $n = 50000$}
\caption{Boxplots of the bootstrap estimation of the tuning parameter $k$.}\label{fig:kEst3d}
\end{center}
\end{figure}

\begin{figure}[ht!]
\begin{center}
\includegraphics[height=6cm,width=7cm]{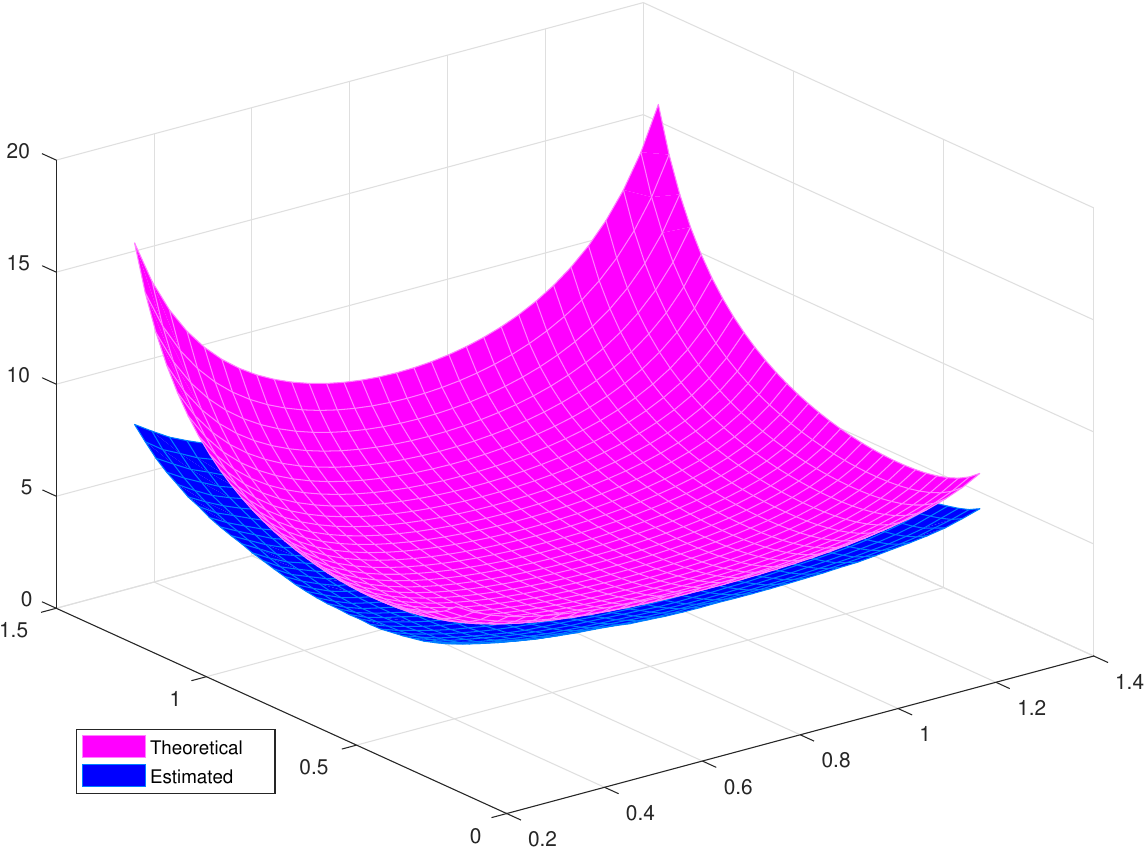}
\includegraphics[height=6cm,width=7cm]{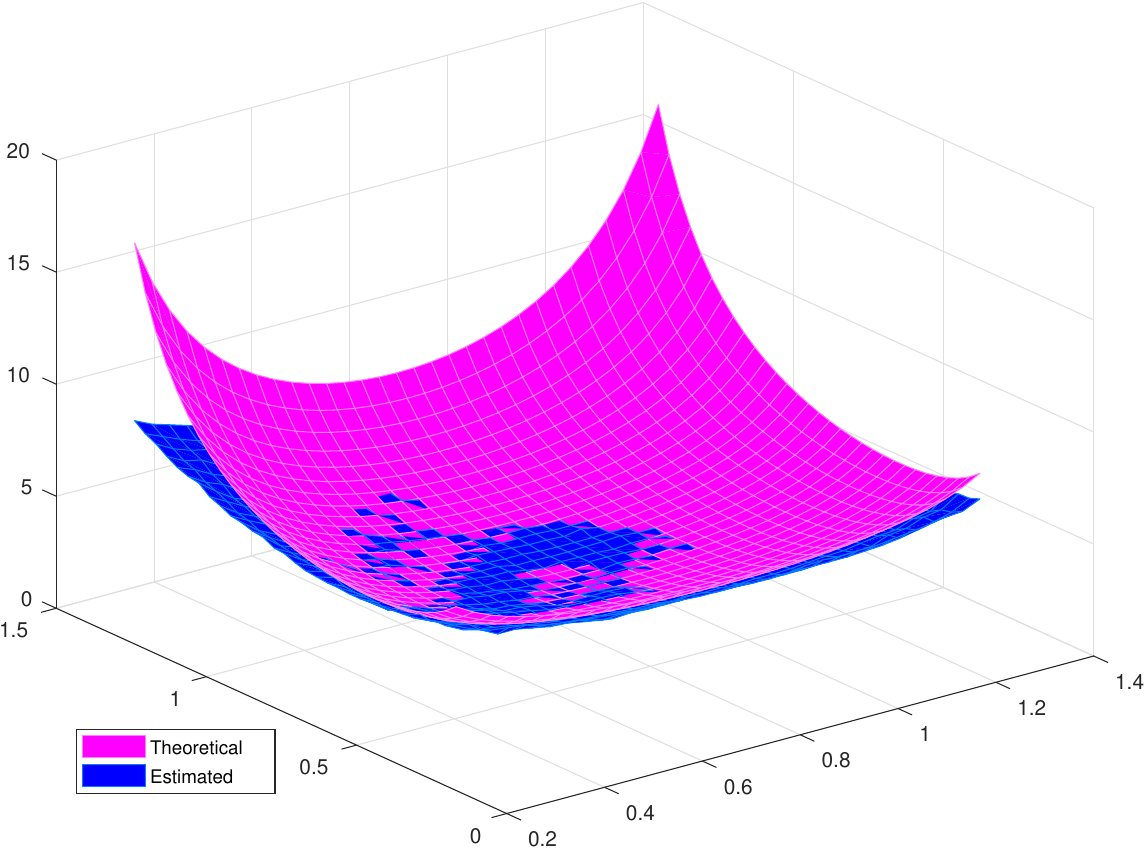}
\vspace{-0.25cm}
\centerline{(A)\;$n=500$\hspace{4cm} (B)\;$n=50000$}
\caption{Theoretical and estimated curves $\rho_{\mathbf{u}}(\boldsymbol\theta)$, with $\mathbf{u}=FPC=(0.8417,0.4202,-0.3392)$. }\label{fig:RhoEst3d}
\end{center}
\end{figure}

\begin{figure}[ht!]
\begin{center}
\includegraphics[height=6cm,width=7cm]{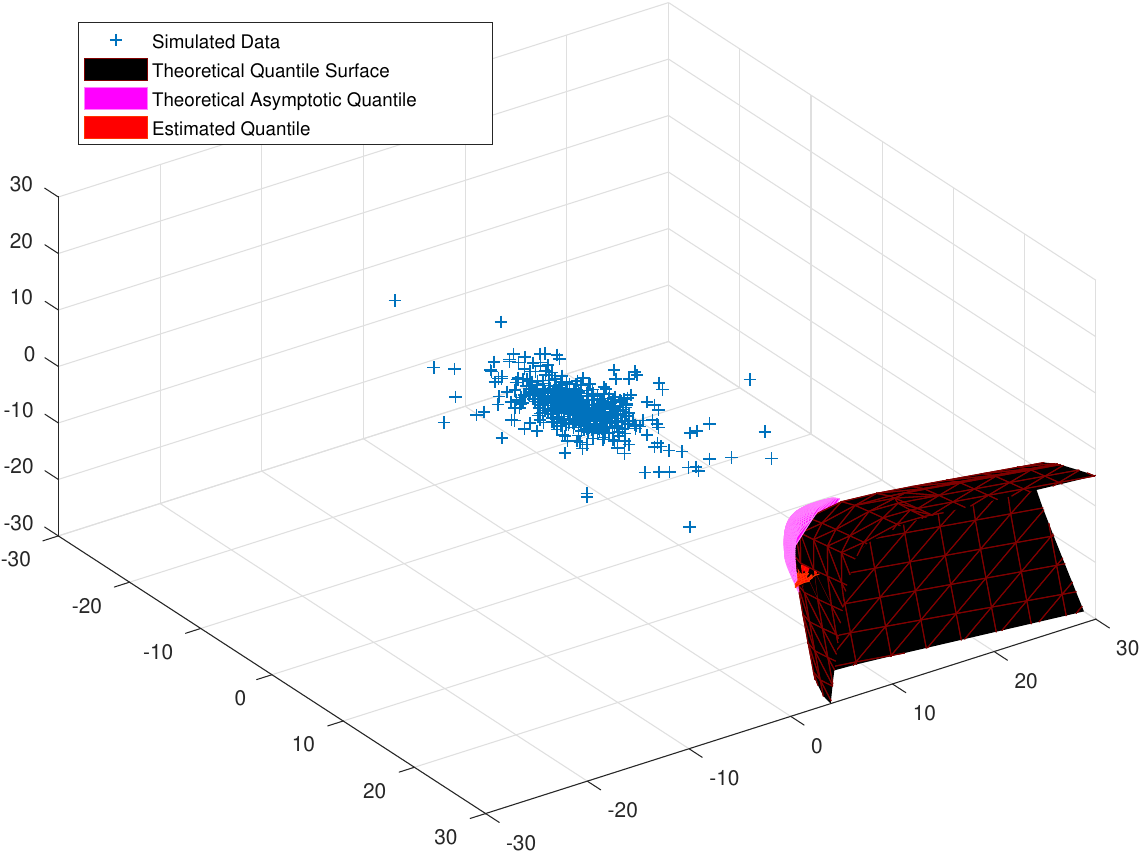}
\includegraphics[height=6cm,width=7cm]{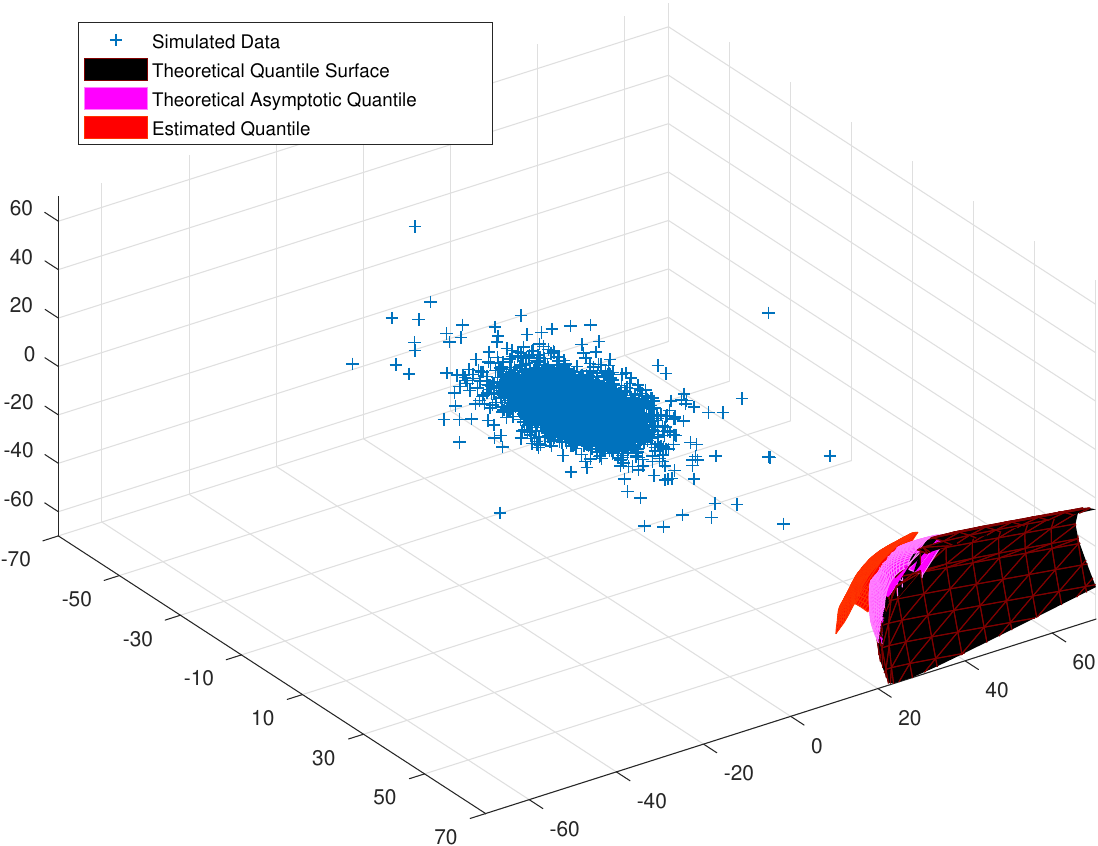}
\vspace{-0.25cm}
\centerline{(A)\;$n=500$\hspace{4cm} (B)\;$n=50000$}
\caption{Estimations for $\mathcal{Q}_{\mathbf{X}}(1/n,FPC)$.}\label{fig:estQFin3d}
\end{center}
\end{figure}

Then, we proceed to illustrate the results through the bootstrap procedure from Figure \ref{fig:kEst3d} to Figure \ref{fig:estQFin3d}, but just in the $FPC$ direction, ($\mathbf{u}=(0.8417,0.4202,-0.3392)$)  and without the construction of the point-wise confidence bands. However, in Figure \ref{fig:kEst3d} we include Monte Carlo replications to describe the distribution of the tuning parameter $k(n)$ selected by bootstrap method. In this trivariate example we consider $n = 500$ and $n = 50000$.

Figure \ref{fig:RhoEst3d} displays the behavior of $\rho_{\mathbf{u}}(\boldsymbol\theta)$ with  $\boldsymbol\theta$ described in terms of the $(d-1)$ angles of its polar parameterization.  The estimation is accurate in general, but specially in the central part of the parametric domain of $\boldsymbol\theta$, which is very important to describe properly the behavior of the directional quantile because it represents the region with maximum curvature in the quantile surface. Figure \ref{fig:estQFin3d} displays the final estimation of $\mathcal{Q}_{\mathbf{X}}(1/n,FPC)$.

\begin{figure}[ht!]
\begin{center}
\includegraphics[height=4.5cm,width=6cm]{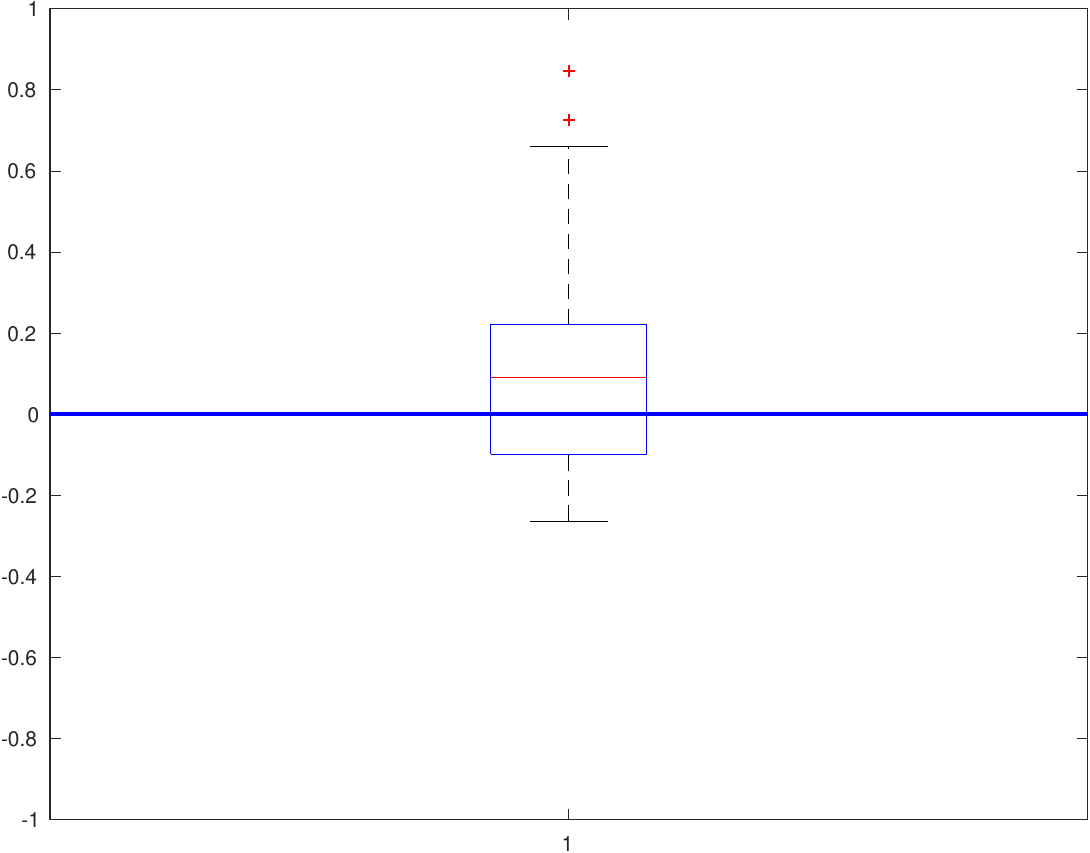}
\includegraphics[height=4.5cm,width=6cm]{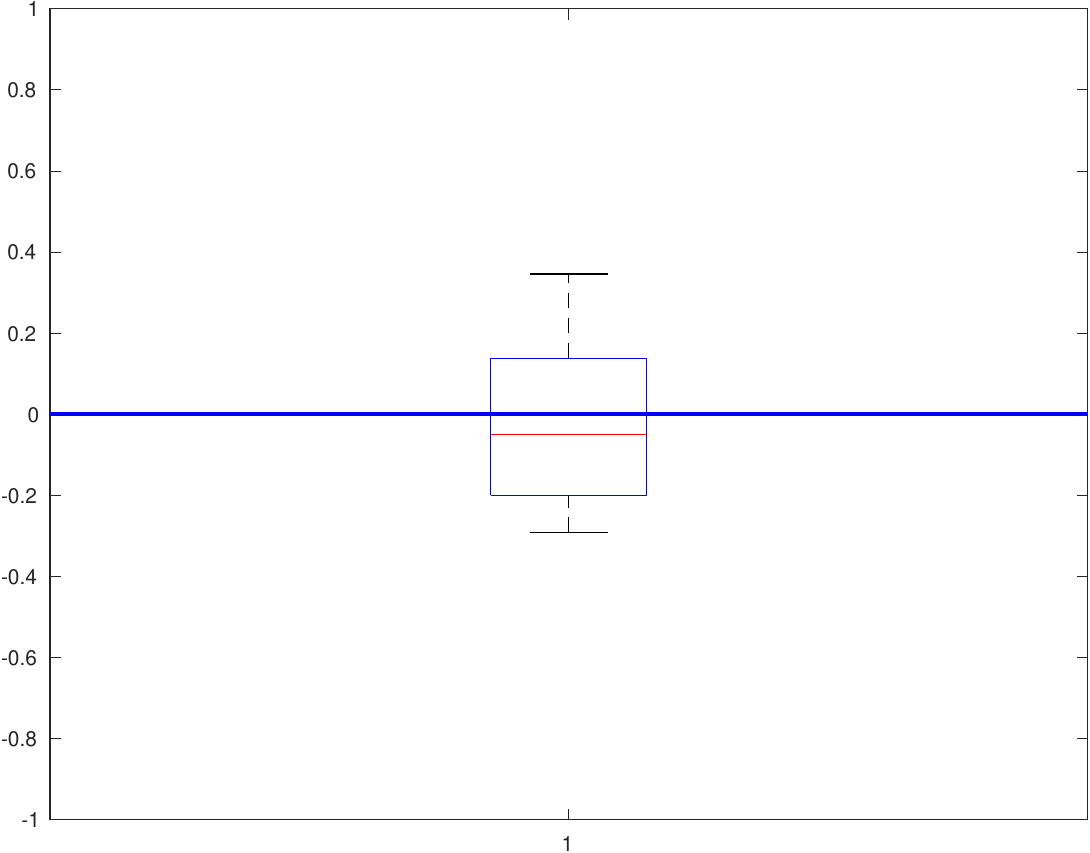}
\vspace{-0.25cm}
\centerline{(A)\;$n=500$\hspace{4cm} (B)\;$n=50000$}
\caption{Relative error ($RE$)  in $\mathcal{Q}_{\mathbf{X}}(1/n,FPC)$ in the point $\boldsymbol\theta = (1/\sqrt{d},\ldots,1/\sqrt{d})$.}\label{fig:QEstPerf}
\end{center}
\end{figure}

The visual performance is quite accurate. We introduce in the following a measurement to assess the quality of these results, i.e.,  the relative error between estimated and asymptotic theoretical quantiles in the point of maximum convexity. By using \eqref{eq:xApprox} and \eqref{eq:xEst}, it can be written as
\begin{equation*}\label{eq:errorEst}
RE = sign(||\hat{\mathbf{x}}_{\mathbf{u}}(\alpha,\boldsymbol\theta, n/k)||-||\tilde{\mathbf{x}}_{\mathbf{u}}(\alpha,\boldsymbol\theta)||)\frac{||\hat{\mathbf{x}}_{\mathbf{u}}(\alpha,\boldsymbol\theta, n/k) - \tilde{\mathbf{x}}_{\mathbf{u}}(\alpha,\boldsymbol\theta)||}{||\tilde{\mathbf{x}}_{\mathbf{u}}(\alpha,\boldsymbol\theta)||},
\end{equation*}
where $\boldsymbol\theta = (1/\sqrt{d},\ldots,1/\sqrt{d})$. Box-plots of $RE$ for the two considered sample sizes are displayed in Figure \ref{fig:QEstPerf}.

\section{Application to financial real data set}\label{sec:realCase}

When the object of study is a financial portfolio, the relevance of analysis in the direction of the vector of weights of investment was pointed out in \cite{torres1st}. In this section, the aim is to highlight that the methodology presented in this paper offers an alternative for decision making when investment allocation and particular management criteria are considered. Therefore, we summarize a real case analyzed previously in \cite{yihe}, describing the main differences between a general analysis of risks through close trimming contours and the specific directional analysis proposed in this paper.

\cite{yihe}  analyzed the daily market return of a portfolio composed of three international indices from July 2nd, 2001 to June 29th, 2007. The indices are the S$\&$P 500 index from USA, the FTSE 100 index from UK and the Nikkei 225 index from Japan. The data contains 1564 observations and it is well-known from the financial literature that stocks returns usually reject the serial independence. Hence, one cannot work with the raw data since the assumption of \emph{i.i.d.}  observations may be inappropriate. \cite{yihe}  filtered the data to solve this issue. Each time series of market returns was modeled by an exponential
 GARCH(1,1) and fitted the parameters by maximizing the quasi-likelihood to obtain the filtered returns, also called \textit{innovations}, which were modeled by a $t-$distribution.

\begin{figure}[ht!]
\begin{center}
\includegraphics[height=6cm,width=7cm]{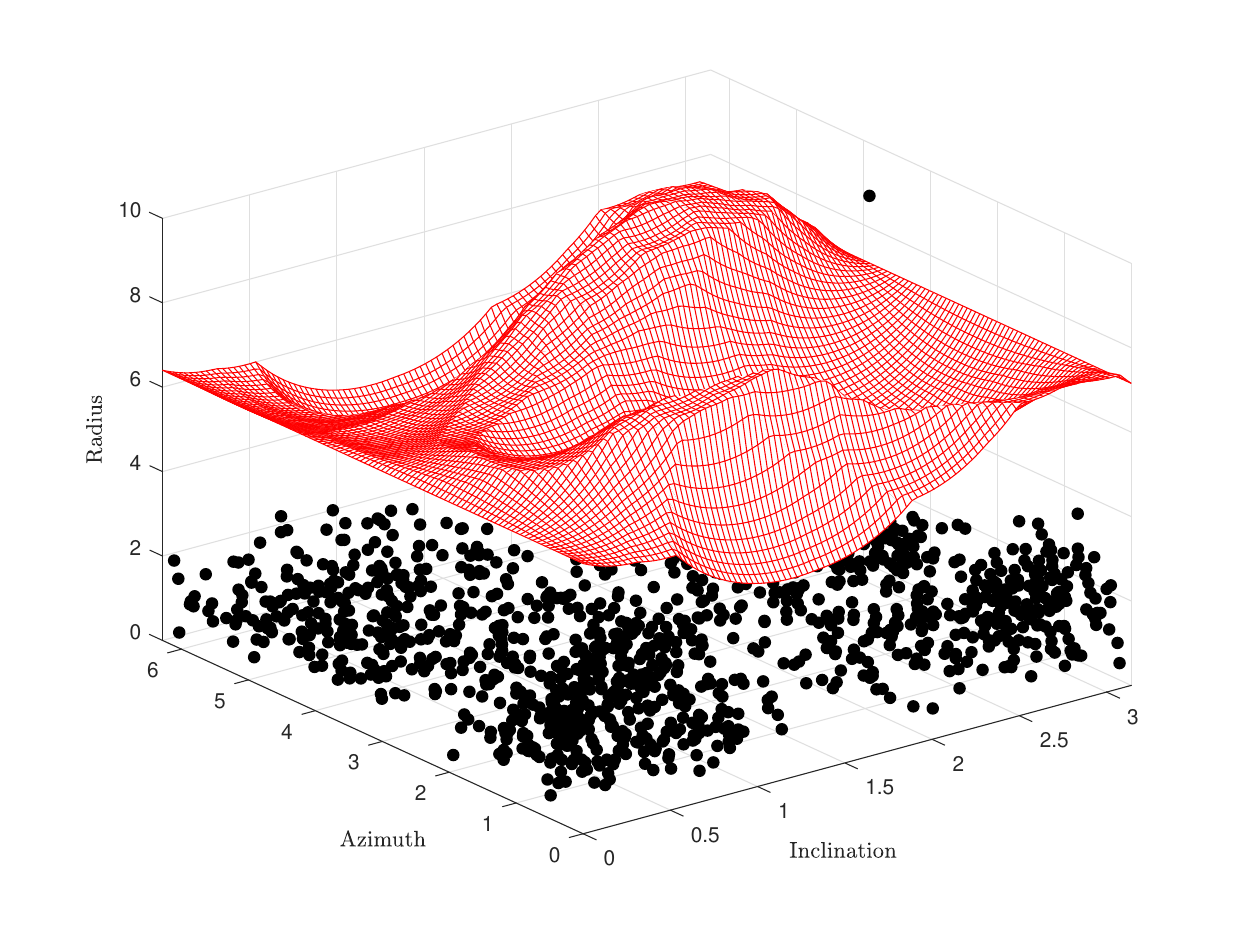}
\vspace{-0.25cm}
\caption{Outlier criteria through Tukey depth trimming for $\alpha = 1/10000$ from \cite{yihe}.}\label{fig:yihe}
\end{center}
\end{figure}

As we pointed out in Section \ref{sec:intro}, methods based on depth and density contours inherently consider the whole set of directions, which provides an overall analysis. However, an analysis considering particular criteria or manager preferences is outside of the aim of those methods. \cite{yihe} used the Tukey depth to build \textit{out-sample} trimming contours for the \textit{innovations} of these  three indices  and they suggested to consider the big loss in the US market on February 27th, 2007 as an outlier by considering its innovation far enough based on the high level contour with $\alpha=1/10000$. Figure \ref{fig:yihe} displays the results in spherical coordinates to support their claim.

Thus, the region above the surface in Figure \ref{fig:yihe} accumulates an approximated probability of $1/10000$ in its global analysis. However, the directional approach concentrates the $\alpha-$level of probability in the set $\{\mathbf{x}\in\mathbb{R}^{d}:\mathbb{P}(\mathfrak{C}_{\mathbf{x}}^{-\mathbf{u}})\geq 1-\alpha\}$, where $\mathbf{u}$ can  incorporate some manager preferences. Then, the QR oriented orthant in Definition \ref{def:orthantQR} can provide an analog rule to identify directional outliers in a similar way as in \cite{yihe}. A naive proposal is to consider the fact that the QR oriented orthant in direction $\mathbf{u}$ divides the $\mathbb{R}^{d}$ space in $2^{d}$ \textit{``disjoint parts''} which leads to the value of $\alpha = 8/10000 = 1/1250$.

For instance, if the criteria of analysis is the portfolio weights of investment and considering $\alpha  = 1/1250$, two examples can be chosen to highlight differences: 1) the naive diversification of the portfolio, i.e., $\mathbf{u} = \mathbf{e}$ and 2) an investment with large participation in the U.S. market, regular in the U.K. market and small in the Japanese market: $\mathbf{u} = (0.6,0.35,0.05)$. The directional analysis is carried out over the filtered losses, i.e., the negative of the \textit{innovations}. As in \cite{yihe}, the filtered losses can be fitted by a multivariate $t-$Student distribution, which allows us to perform the directional approach in twofold:
\begin{enumerate}[1)]
\item A semi-parametric method; that is, the estimation of the parameters of the $t-$model for the \textit{negative innovations} and the calculation of the theoretical directional iso-surfaces for this  model by using tools presented in Section \ref{sec:tExample}.
\item Our full non-parametric method presented in Section \ref{sec:inference},  considering the fact that \cite{yihe} previously tested the multivariate regular variation condition on this data-set by  the method in \cite{einmahl-krajina}.
\end{enumerate}

\begin{figure}[ht!]
\begin{center}
\includegraphics[height=6cm,width=7cm]{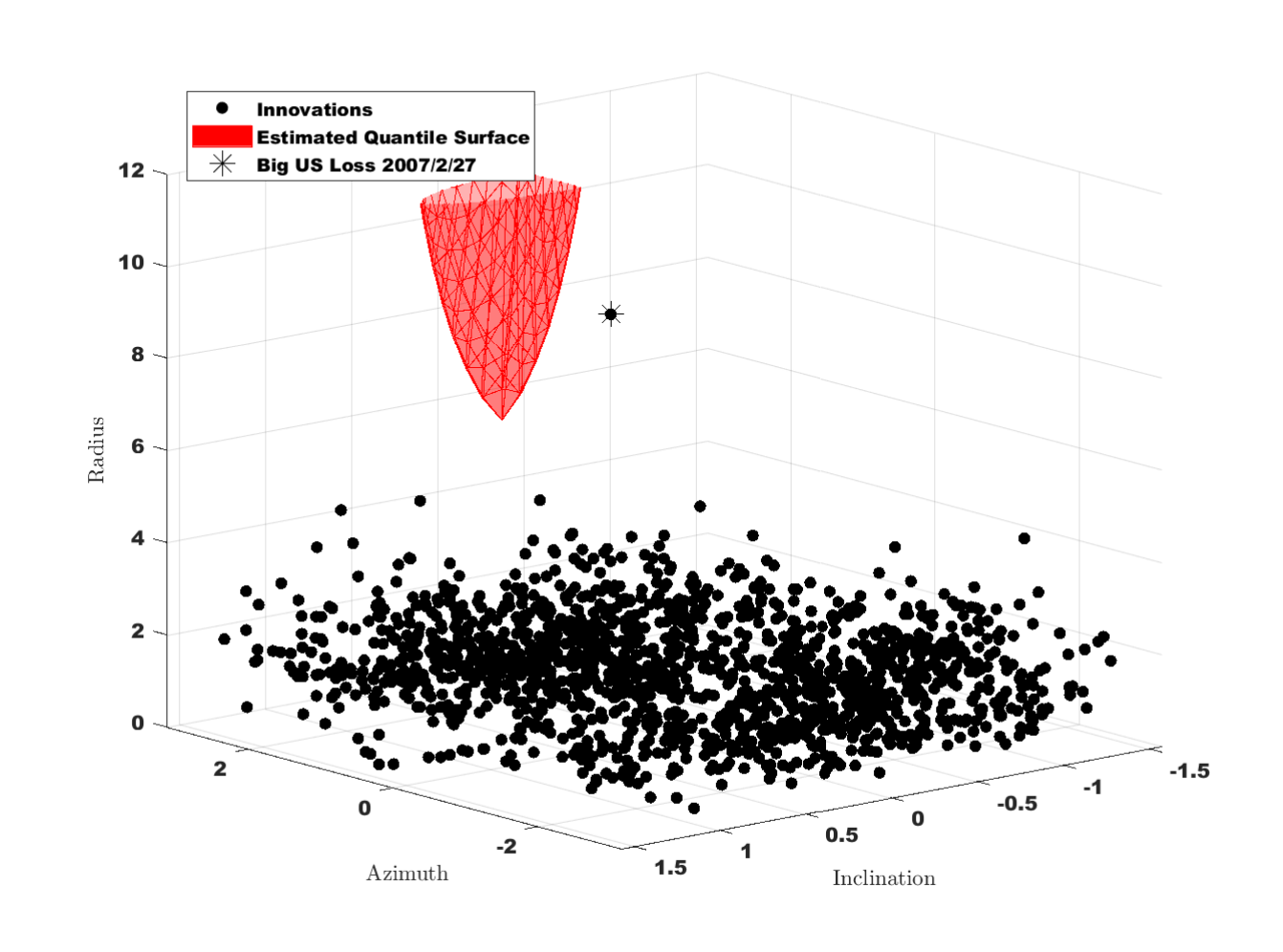}
\includegraphics[height=6cm,width=7cm]{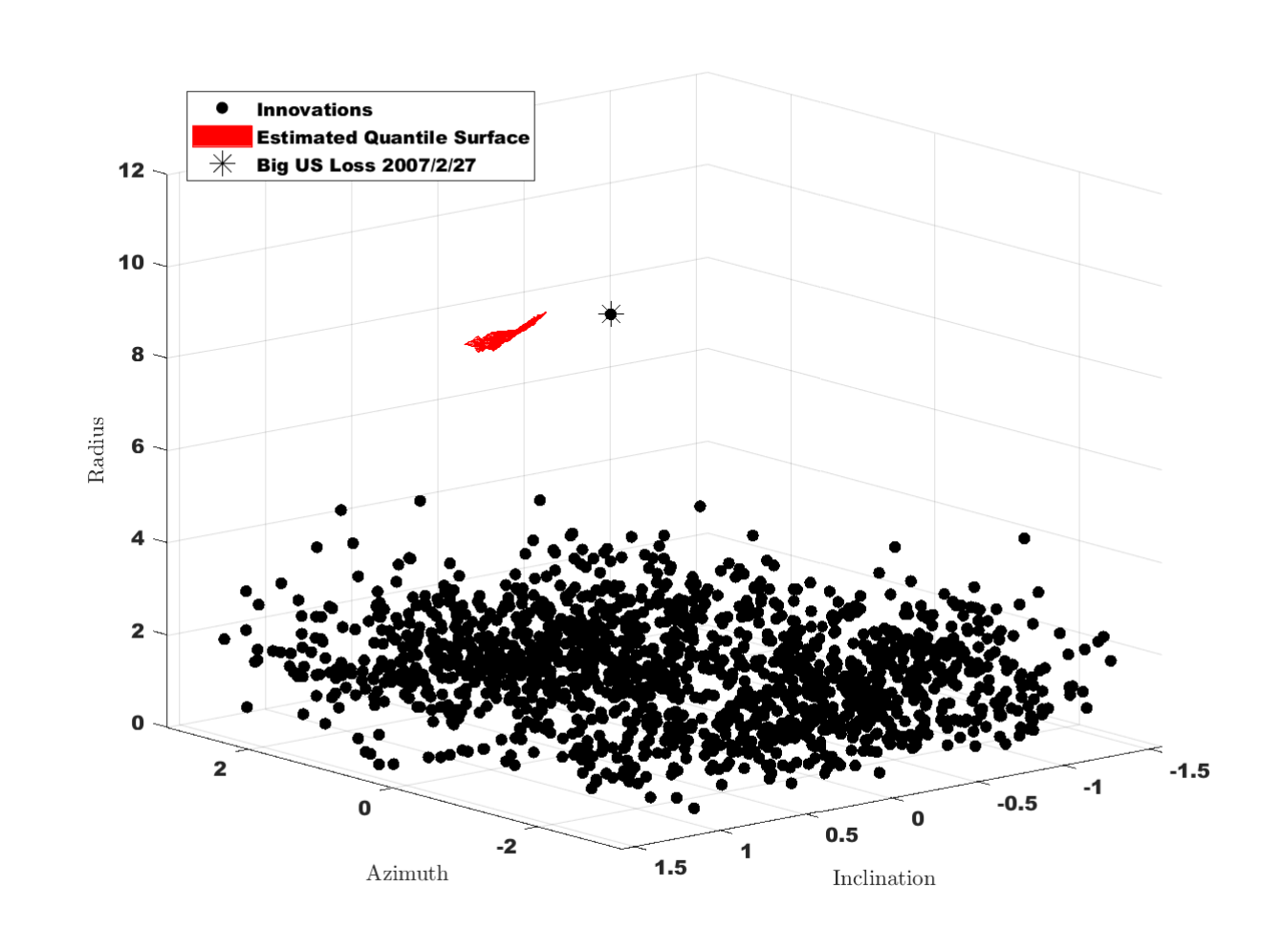}
\vspace{-0.25cm}
\centerline{(A)\;Semi-parametric approach\hspace{2cm} (B)\;Non-parametric approach}
\caption{Directional portfolio criteria, $\mathbf{u}=\mathbf{e}$ and $\alpha = 1/1250$.}\label{fig:resultsEvenInv}
\end{center}
\end{figure}

\begin{figure}[ht!]
\begin{center}
\includegraphics[height=6cm,width=7cm]{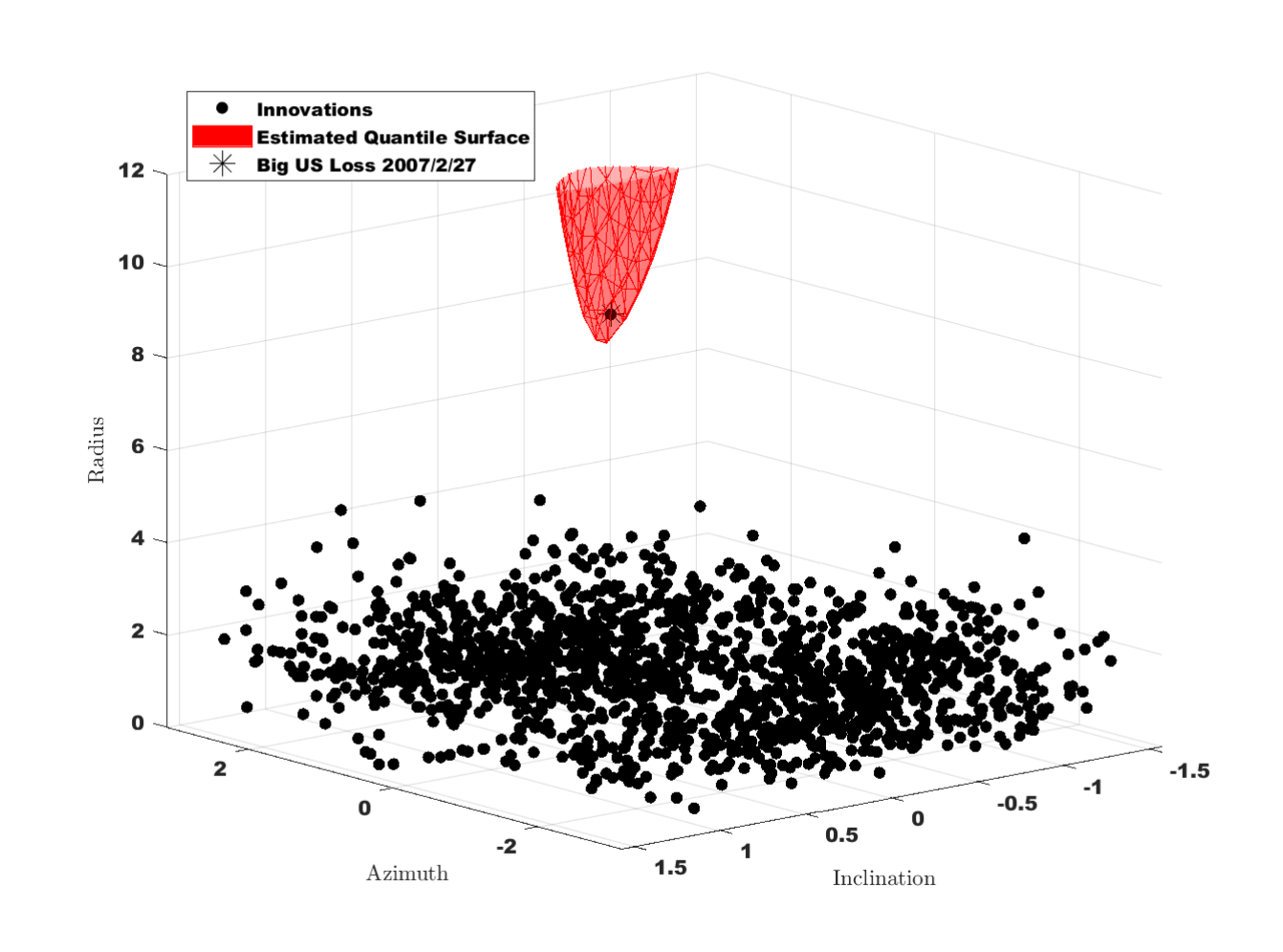}
\includegraphics[height=6cm,width=7cm]{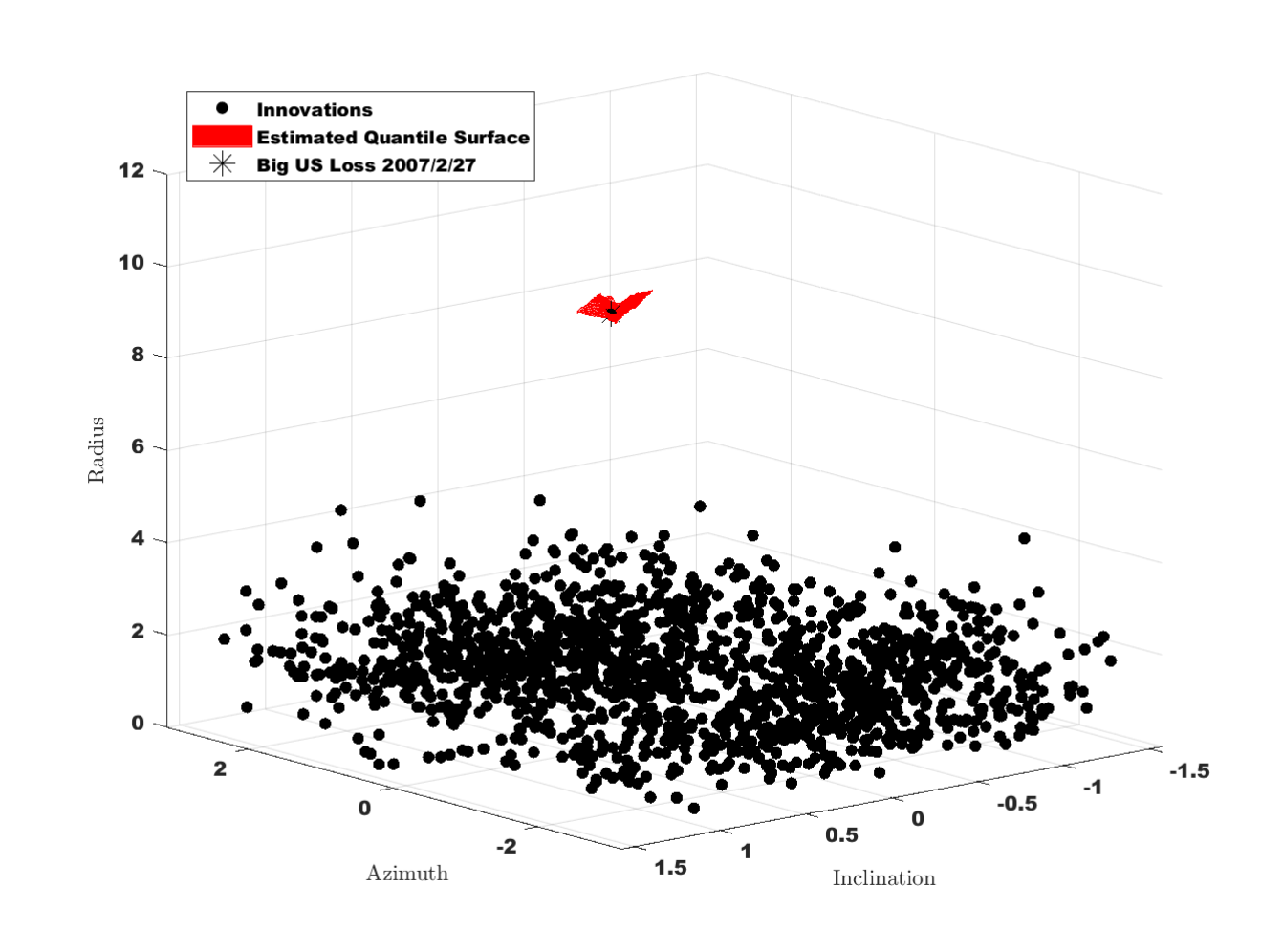}
\vspace{-0.25cm}
\centerline{(A)\;Semi-parametric approach\hspace{2cm} (B)\;Non-parametric approach}
\caption{Directional portfolio criteria, $\mathbf{u}=(0.6,0.35,0.05)$ and $\alpha = 1/1250$.}\label{fig:resultsOddInv}
\end{center}
\end{figure}

In Figure \ref{fig:resultsEvenInv} we focus in the classical direction $\mathbf{e}$. One can see that the big U.S. loss is not identified here as an outlier point because it is not contained in the critical region. This suggests a leverage effect that cannot be underestimated for this particular investment. Conversely,  Figure \ref{fig:resultsOddInv} shows that the so called big U.S. loss is indeed above the critical layer for the investment weights $\mathbf{u} = (0.6,0.35,0.05)$. This  leads to a similar interpretation of outlier to the one provided by \cite{yihe}.

\cite{yihe} commented that ``\textit{Neglecting the joint behaviour can lead to an overestimated diversifiability of risks across international markets and, therefore, underestimation of systematic risk}''. In this sense, we add that the directional approach allows to include external information or manager criteria providing a joint local analysis that could lead to different conclusions than those in an overall joint behavior.  Furthermore, in \cite{yihe} is noted that ``\emph{an outlier in a high dimensional space is not necessarily an outlier in its subspaces with reduced dimensions}''. We also point  out that an outlier in one direction is not necessarily an outlier in the other ones.

\section{Conclusions}\label{sec:conclusions}
The $MEVT$ theory has been extended by the inclusion of the directional framework into it. Also, this paper presented an \textit{out-sample} characterization of those DMQ $\mathcal{Q}_{\mathbf{X}}(\alpha,\mathbf{u})$, recently introduced in \cite{torres1st,torres2nd}. Necessary conditions to ensure the estimation of the \textit{DMQ} at high levels independently of the chosen direction $\mathbf{u}$ were also presented and the proposed estimator integrates different asymptotic results from the univariate and the multivariate extreme value theory through a parameterization in polar coordinates in $\mathbb{R}^{d}$.

We introduced an adapted bootstrap-based method to find an optimal solution for the tuning parameter $k$ in this multivariate framework, joint to a non-parametric method to complete the estimation of the \textit{DMQ}. Finally, asymptotic normality of the estimator was derived.

Based on the multivariate $t-$distribution, illustrations of the estimation procedure in dimensions 2 and 3 are shown. This family of distributions possesses properties such as heavy tails and closure under rotations, which provides a good example for comparing theoretical and estimated solutions. Finally, a real case study identifying directional outliers in the filtered losses  of a financial portfolio is performed. This example suggests that joint local analysis could lead to different conclusions than overall joint behavior. This provides a  wider vision to the fact that neglecting the joint behavior can lead to an overestimated diversifiability of risks across international markets.
%

Future interesting works are to extend the directional framework to the multivariate max-domain of attraction setting, i.e., to relax  Assumption \ref{assum3}. Also, to analyze if there exist improvement of estimation by setting independent optimal tuning parameters, for instance, $k_{\rho}$ for the estimation of $\tilde{\rho}_{\mathbf{u}}$ in Equation \eqref{eq:empLimitDist}. Finally, focusing on applications, it is a task to build a multivariate Value-at-Risk measure in the \textit{out-sample} framework based on \textit{DMQ} to analyze risks in different real case scenarios. And also, it is demanding a definition of return period in a general multivariate framework for environmental problems.

\appendix

\section{Auxiliary results and proofs}\label{sec:proofs}
This section is devoted to the proofs of main results of this paper. Furthermore, different necessary results are introduced below.\smallskip

\begin{Proof}[Proof of Proposition \ref{prop:rotRegVar}]
For each Borel set $B$, we have that $QB := \{QX | X\in B\}$ is a Borel set. Let  denote $\mathbb{P}$ and $\mathbb{P}_{Q}$ as the probability measures of $\mathbf{X}$ and $Q\mathbf{X}$, respectively. For a Borel set $B$, one can write
\[\mathbb{P}[B]=\mathbb{P}_{Q}[QB]\quad \text{or analogously} \quad \mathbb{P}[Q'B]=\mathbb{P}_{Q}[B].\]
Therefore, we obtain that the random vector $Q\mathbf{X}$ is also multivariate regularly varying with tail index $\gamma$ since
\[t\mathbb{P}_{Q}\left[\frac{Q\mathbf{X}}{\phi(t)}\in \cdot\right]\stackrel{v}{\rightarrow} \mu_{Q}(\cdot):=\mu\circ Q'(\cdot).\]
\end{Proof}
\vspace{0.1cm}

\begin{Proof}[Proof of Proposition \ref{prop:rotRegVar2}]
As in  the proof of Proposition \ref{prop:rotRegVar}, we denote $\mathbb{P}$ and $\mathbb{P}_{Q}$  the probability measures of $\mathbf{X}$ and $Q\mathbf{X}$, respectively. Then, we get for any relatively compact rectangle $B$ that,
\[\frac{t\mathbb{P}_{Q}\left[\frac{Q\mathbf{X}}{\phi(t)}\in B\right]-\mu_{Q}(B)}{\Lambda(\phi(t))} = \frac{t\mathbb{P}\left[\frac{\mathbf{X}}{\phi(t)}\in Q'B\right]-\mu(Q'B)}{\Lambda(\phi(t))}.\]
Hence,
\[\frac{t\mathbb{P}_{Q}\left[\frac{Q\mathbf{X}}{\phi(t)}\in B\right]-\mu_{Q}(B)}{\Lambda(\phi(t))}\rightarrow \psi_{Q}(B) := \psi\circ Q'(B).\]
\end{Proof}
\vspace{0.1cm}

For the sake of readability, we now introduce the directional multivariate versions of the four lemmas, Lemma 2.1 to Lemma 2.4 in \cite{dehaan1995} without proof. These four lemmas will be useful to prove Proposition \ref{prop:normality} below.

\begin{Lemma}\label{lemma1}
If $-\ln G_{\mathbf{u}}$ has continuous first-order derivatives $(-\ln G_{\mathbf{u}})_{i},\, i=1, ..., d$, then
\[\sqrt{k}\left(\hat{\rho}_{\mathbf{u}}(\boldsymbol\theta)\,-\, \tilde{\rho}_{\mathbf{u}}(\boldsymbol\theta)\right)\]
converges to  \,  $V_{\mathbf{u}}\left(\frac{\theta_{j}^{\gamma}-1}{\gamma};\, j=1, ..., d\right) + \sum_{i=1}^{d}(-\ln G_{\mathbf{u}})_{i}\left(\frac{\theta_{j}^{\gamma}-1}{\gamma};\, j=1, ..., d\right)\left[\int_{1}^{\theta_{i}}(\ln t)t^{\gamma-1}dt\right]\Gamma_{\mathbf{u},i}.$
\end{Lemma}
\vspace{0.1cm}

\begin{Lemma}\label{lemma2}
Under the conditions of Proposition \ref{prop:normality},
\[\sqrt{k}\left(\frac{\hat{x}_{\mathbf{u},j}(\alpha,\mathbf{e},\boldsymbol\theta,n/k)\, -\, \tilde{x}_{\mathbf{u},j}(\alpha, \boldsymbol\theta)}{\hat{a}_{\mathbf{u},j}(n/k)\int_{1}^{s_{n}}t^{\hat{\gamma}_{j}-1}(\log t)dt}\right)\]
converges in distribution to \,
$\left(\rho_{\mathbf{u}}(\boldsymbol\theta)\,\theta_{j}\right)^{\gamma}\Gamma_{\mathbf{u},j}$, for all $j=1, ..., d$.
\end{Lemma}

\vspace{0.1cm}
\begin{Lemma}\label{lemma3}
Under the conditions of Proposition \ref{prop:normality},
\[\lim_{n\rightarrow\infty}\sqrt{k}\left(\rho_{\mathbf{u}}(\boldsymbol\theta)\, -\, \tilde{\rho}_{\mathbf{u}}(\boldsymbol\theta)\right)= 0
\,\,  \mbox{ locally uniformly. } \]
\end{Lemma}
\vspace{0.1cm}

\begin{Lemma}\label{lemma4}
Let  $\psi_{\mathbf{u},j}(x_{j}) := \psi_{\mathbf{u}}\left([-\infty,\infty]\times\cdots\times [x_{j}, \infty]\times\cdots\times [-\infty,\infty]\right)\quad j=1,\ldots, d$.
Then, under the conditions of Proposition \ref{prop:normality},
\[\lim_{n\rightarrow\infty}\frac{\sqrt{k}\left(\tilde{x}_{\mathbf{u},j}(\alpha, \boldsymbol\theta)\, -\, x_{\mathbf{u},j}(\alpha, \boldsymbol\theta)\right)}{\hat{a}_{\mathbf{u},j}(n/k)s_{n}^{\hat{\gamma}_{j}+1}\,\psi_{\mathbf{u},j}
\left(\left(s_{n}^{\hat{\gamma}_{j}}-1\right)/\hat{\gamma}_{j}\right)}=0, \,\, \mbox{ locally uniformly,  for all }   j=1, ..., d.\]
\end{Lemma}

The proofs of these lemmas work in a similar way as in \cite{dehaan1995}  considering the arrangements due to the directional multivariate framework, then they are omitted here. Now, by using Lemmas \ref{lemma1}-\ref{lemma4}, one can prove  the main Proposition \ref{prop:normality}.\smallskip

\begin{Proof}[Proof of Proposition \ref{prop:normality}]
Lemma \ref{lemma1} proves asymptotic convergence of the standardized difference $\hat{\rho}_{\mathbf{u}}(\boldsymbol\theta)\,-\, \tilde{\rho}_{\mathbf{u}}(\boldsymbol\theta)$. This implies asymptotic normality of the standardized difference $\hat{x}(\alpha,\mathbf{e},\boldsymbol\theta,n/k)\, -\, \tilde{x}_{\mathbf{u},j}(\alpha, \boldsymbol\theta)$ in Lemma \ref{lemma2}.

Also, Lemma \ref{lemma3} proves the convergence to zero of the standardized difference $\rho_{\mathbf{u}}(\boldsymbol\theta)\, -\, \tilde{\rho}_{\mathbf{u}}(\boldsymbol\theta)$, which helps to prove Lemma \ref{lemma4} where the convergence to zero of the standardized difference $\tilde{x}_{\mathbf{u},j}(\alpha, \boldsymbol\theta)\, -\, x_{\mathbf{u},j}(\alpha, \boldsymbol\theta)$ is given. Then, by using the asymptotic normality of the standardized difference between the approximation and the estimation in Lemma \ref{lemma2} and the convergence to zero of the standardized difference between the true elements and its approximations in Lemma \ref{lemma4}, the result in Proposition \ref{prop:normality} is achieved.
\end{Proof}
\vspace{0.3cm}

\vspace{0.3cm}
We recall below an useful result for elliptical distribution (see   Lemma 3.1 in \cite{hult}).
\begin{Lemma}\label{lemma:elliptic}
If $\mathbf{X}$ has an elliptical distribution and decomposition given by
\[\mathbf{X} \overset{d}{=} \boldsymbol\mu + \Sigma^{1/2}\,r\,\mathbf{Z},\]
where $r$ is a random variable independent from the random vector $\mathbf{Z}$, which is uniformly distributed in the unit circle of dimension $d$, $\boldsymbol\mu$ a location parameter and $\Sigma$ a matrix indicating scale. Then, for any orthogonal matrix $Q$, $Q\mathbf{X}$ has an elliptical distribution with associated decomposition given by
\[Q\mathbf{X} \overset{d}{=} Q \boldsymbol\mu + Q\Sigma^{1/2}\,r\,\mathbf{Z}.\]
Moreover, its marginals are the associated univariate elliptical distributions with parameters of location and scale given by $Q \boldsymbol\mu$ and $Q\Sigma Q'$.
\end{Lemma}

Now we summarize the result from \cite{niko}, where the theoretical stable tail dependence function for a multivariate $t-$distribution is obtained.
The interested reader is also referred to  \cite{opitz}.
\begin{Theorem}[\cite{niko}, Theorem 2.3]\label{remarkNiko}
The theoretical tail function of $T_{0,\Sigma,\nu}^{d}(\cdot)$, a $d-$dimensional $t-$distribution with d.f. $\nu$, location parameter $\boldsymbol\mu = 0$ and scale parameter $\Sigma$, is given by,
\begin{small}
\begin{equation*}\label{eq:tailT}
\begin{aligned}
-\ln\left(G\left(\frac{\mathbf{z}^{\gamma}-\mathbf{1}}{\gamma}\right)\right) = \sum_{j=1}^{d}z_{j}^{-1} \,T_{0,Q_{j},\nu+1}^{d-1}\left(\sqrt{\frac{\nu+1}{1-r_{i,j}^{2}}}\left[\left(\frac{z_{i}}{z_{j}}\right)^{1/\nu}-r_{i,j}\right];\ \ i\neq j\right),
\end{aligned}
\end{equation*}
\end{small}

where $r_{i,j}$ are the correlations between the components $i,\, j$, $T_{0,Q_{j},\nu+1}^{d-1}(\cdot)$ is a $t-$distribution in dimension $d-1$ (removing the $j-$component), with d.f. $\nu+1$, location parameter $\boldsymbol\mu = 0$ and scale parameter given by,
\[Q_{j}=\begin{bmatrix}
1 & \cdots & r_{1,j-1;j} & r_{1,j+1;j} & \cdots & r_{1,d;j}\\
\vdots & \ddots & \vdots & \vdots & \cdots & \vdots \\
r_{j-1,1;j} & \vdots & 1 & r_{j-1,j+1;j} & \cdots & r_{j-1,d;j}\\
r_{j+1,1;j} & \vdots & r_{j+1,j-1;j} & 1 & \cdots & r_{j+1,d;j}\\
\vdots & \cdots & \vdots & \vdots & \ddots & \vdots \\
r_{d,1;j} & \vdots & r_{d,j-1;j} & r_{d,j+1;j} & \cdots & 1\\
\end{bmatrix},\]
where $r_{i,l;j}=\frac{r_{il}-r_{ij}r_{lj}}{\sqrt{1-r_{ij}^{2}}\sqrt{1-r_{lj}^{2}}}$, for $i,\, l \neq j$.
\end{Theorem}

\section*{Acknowledgments}

This research was partially supported by a Spanish Ministry of Economy and Competitiveness grant ECO2015-66593-P, the French project LEFE-MANU MULTIRISK and by Consejería de Educación de la Junta de Castilla y León and FEDER funds (ref. VA005P17).  

\end{document}